\documentclass{iopart}
\usepackage{natbib}
\usepackage{graphicx}
\usepackage{bbm}
\usepackage{aas_macros}
\usepackage{amsmath}
\usepackage{amssymb}

\newif\ifpdf
\ifx\pdfoutput\undefined
\pdffalse
\else
\pdfoutput=1
\pdftrue
\fi
\ifpdf
\usepackage{graphicx}
\usepackage{epstopdf}
\else
\usepackage{graphicx}
\fi

\usepackage{color}

\newcommand{\sun}{\ensuremath{_\odot}}
\newcommand{\dd}{\,\textrm{d}}

\newcommand{\pcc}{\textrm{cm}\ensuremath{^{-3}}}

\newcommand{\muG}{\ensuremath{\mu}\textrm{G}}

\newcommand{\degree}{\ensuremath{^\circ}}

\begin{document}

\title[Constraining models of the large scale Galactic magnetic field]{Constraining models of the large scale Galactic magnetic field with WMAP5 polarization data and extragalactic rotation measure sources}

\author{Ronnie Jansson$^1$, Glennys R. Farrar$^1$, Andre H. Waelkens$^2$, Torsten A. En\ss lin$^2$}

\address{$^1$Center for Cosmology and Particle Physics,
and Department of Physics\\ New York University, NY, NY 10003, USA\\}

\address{$^2$Max-Planck-Institute f\"ur Astrophysik, Karl Schwarzschild-Str. 1, 85741 Garching, Germany}

\begin{abstract}
We introduce a method to quantify the quality-of-fit between data and observables depending on the large scale Galactic magnetic field. We combine WMAP5 polarized synchrotron data and rotation measures of extragalactic sources in a joint analysis to obtain best fit parameters and confidence levels for GMF models common in the literature. None of the existing models provide a good fit in both the disk and halo regions, and in many instances best-fit parameters are quite different than the original values. We note that probing a very large parameter space is necessary to avoid false likelihood maxima. The thermal and relativistic electron densities are critical for determining the GMF from the observables but they are not well constrained. We show that some characteristics of the electron densities can already be constrained  using our method and with future data it may be possible to carry out a self-consistent analysis in which models of the GMF and electron densities are simultaneously optimized.
\end{abstract}

\section{Introduction}
Among the probes of Galactic magnetism, Faraday rotation measures (RMs) and synchrotron radiation are among the best-suited for studying the large scale Galactic magnetic field. Other probes are either mostly applicable to the nearby part of the Galaxy (starlight polarization) or mainly used to study small scale structures like dense clouds (the Zeeman effect). While synchrotron radiation has been used extensively for studying external galaxies, Faraday rotation measure has been the method of choice for probing the magnetic field of our own Galaxy \citep[e.g.,][]{Han:2006, Brown:2007}. To most simply probe the large scale Galactic magnetic field (GMF) with synchrotron radiation, a full-sky polarization survey is needed, at frequencies where Faraday depolarization effects are negligible and where other sources of polarized radiation (e.g., dust) are either negligible compared to synchrotron radiation or possible to exclude from the data. The 22 GHz K band of the Wilkinson Microwave Anisotropy Probe (WMAP) provides such a data set. 

In this paper we constrain models of the large scale Galactic magnetic field common in the literature by a joint analysis of the polarization data in the WMAP5 K band and rotation measures of extragalactic radio sources. The synergistic advantage of combining these two data sets goes beyond the benefit of achieving a larger number of data points; the two observables probe mutually orthogonal components (perpendicular and parallel to line-of-sight) of the magnetic field.  We calculate the $\chi^2$ for a particular choice of GMF model and parameter choice, and use a Markov Chain Monte Carlo algorithm to optimize the parameters of each field under consideration. Thus we can quantitatively measure the capability of different models to reproduce the observed data. 

The content of this paper is structured as follows: we briefly review the theory of synchrotron radiation and Faraday rotation and their connection with magnetic fields in section \ref{observables}. Section \ref{method} explains our general approach to constrain models of the GMF. Section \ref{data} describes the two data sets used in the analysis. Section \ref{models} details the various 3D models of the magnetized interstellar medium we have studied. Section \ref{results} give our results and a discussion, and is followed by a summary and conclusions.

\section{Observables and input}\label{observables}

\subsection{Faraday rotation measures}\label{obs_egs}

As first shown by Faraday, the polarization angle of an electromagnetic wave propagating through a magneto-ionized medium rotates in proportion to its wavelength squared, 
\begin{equation}
\Delta \theta = \textrm{RM}\,\lambda^2.
\end{equation}
The \emph{rotation measure}, or RM, in units of rad $\textrm{m}\ensuremath{^{-2}}$, is
\begin{equation}
\textrm{RM}\simeq 0.81\int_0^{L}\left(\frac{n_e}{\pcc} \right) \left(\frac{B_\parallel}{\muG} \right) \left(\frac{\dd z}{\textrm{pc}} \right),
\end{equation}
where $n_e$ is the total density of ionized electrons, which is dominated by the \emph{thermal} electron density. We follow the convention where a magnetic field pointing towards the observer causes a positive rotation measure.


\subsection{Synchrotron radiation}\label{obs_synch}

Relativistic electrons spiralling along magnetic field lines radiate synchrotron radiation \citep{Rybicki:1986}. For a power law distribution of relativistic electrons -- commonly called ``cosmic ray'' electrons, $n_{cre}$ -- characterized by a spectral index $s$,
\begin{equation}
n_{cre}(E)dE \propto n_{cre, 0} E^{-s}dE,
\end{equation}
where $n_{cre, 0}$ is a normalization factor, the synchrotron emissivity is 
\begin{equation}
j_\nu\propto n_{cre, 0}B_\perp^{\frac{1+s}{2}}\nu^{\frac{1-s}{2}}.
\end{equation}
For a regular magnetic field and the above relativistic electron distribution, the emitted synchrotron radiation has a large degree of linear polarization, around 75\% for a power law distribution of electron with spectral index $s=3$. Observationally, the \emph{percentage} of polarization is typically much lower due to depolarizing effects such as Faraday depolarization and the presence of turbulent or otherwise irregular magnetic fields which depolarize the radiation through line-of-sight averaging. In this paper we are only interested in the large scale regular field, and will thus exclusively use the polarized components of the observed synchrotron radiation.

\subsection{Input}

As is clear from sections \S \ref{obs_synch} and \S \ref{obs_egs}, the observables we are using are not direct measures of the Galactic magnetic field itself, but convolutions of the various components of the magnetized interstellar medium ($\vec{B}, n_e$ and $n_{cre}$). The ideal way to constrain models of the GMF would thus be to simultaneously and self-consistently  model the thermal and relativistic electron distributions together with the magnetic field. This is beyond the scope of the present work, but should become feasible when even larger data sets become available. Instead we use 3D models of $n_e$ and $n_{cre}$ present in the literature (detailed in section \ref{models}) and do not vary their parameters, except in a separate study of the magnetic field scale height's dependence on the scale height of the thermal electron  distribution in section \ref{scaleheights}.

\section{Method}\label{method}

\begin{figure}
 \begin{center}
\includegraphics[width=\linewidth]{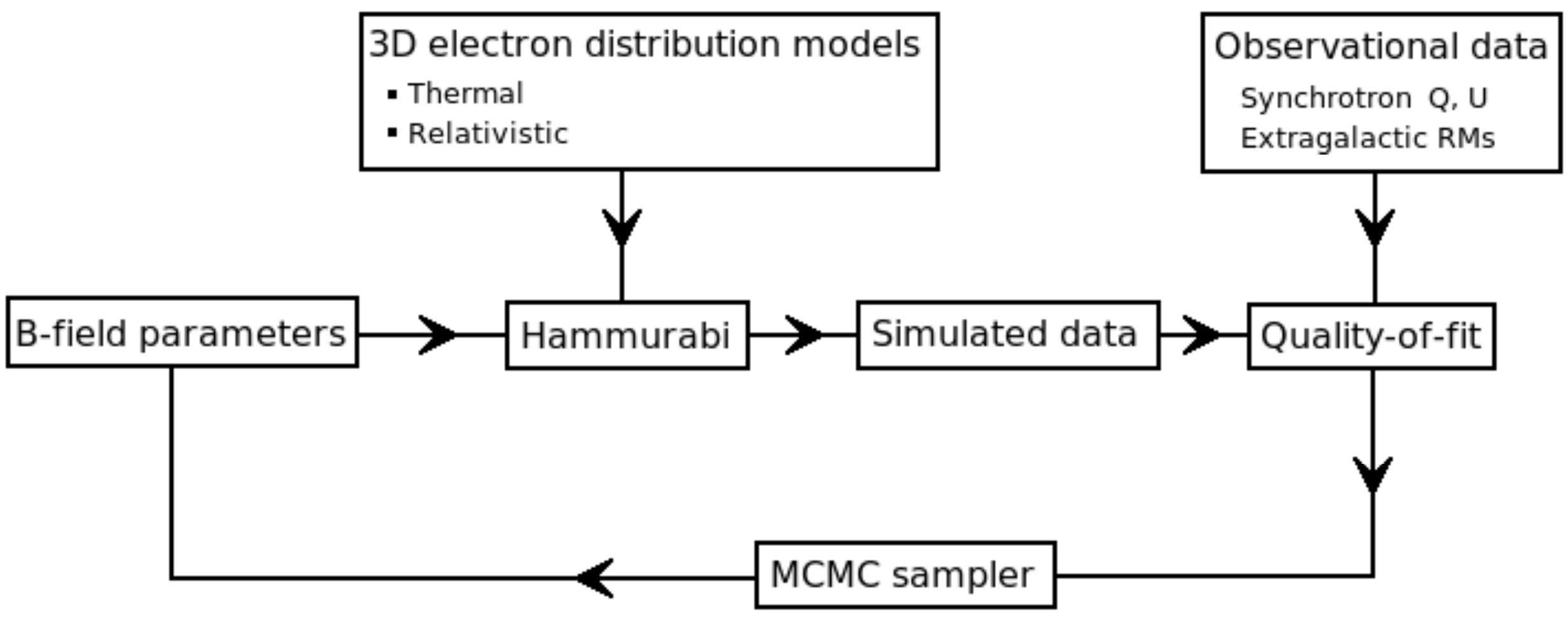}
 \end{center}
\caption{Overview of the implemented analysis.}\label{imp}
\end{figure}

Our strategy to model the GMF is straightforward. As illustrated in figure \ref{imp}, starting from 3D models of the distribution of relativistic electrons ($n_{cre}$), thermal electrons ($n_e$) and a model of the large scale GMF parametrized by a set of numbers $\vec{p}_0$ we produce simulated data sets of polarized synchrotron radiation and Faraday rotation measures. Generating the simulated data is done using the \textsc{Hammurabi} code \citep{Waelkens:2008}. We compare the simulated data sets to the observational data and calculate $\chi^2$, as a measure of the quality-of-fit. This value together with $\vec{p}_0$ is passed to a Markov Chain Monte Carlo (MCMC) sampler that generates a new set of GMF parameters, $\vec{p}_1$, as described in section \ref{paraest}. This scheme is then iterated until we obtain a Markov Chain that has sufficiently sampled the parameter space. In this way we find the best fit parameters and confidence levels for each GMF model, and can compare and quantify the capability of different models to reproduce the observed data.

\subsection{Quality-of-fit}

We define $\chi^2$ in the usual way,
\begin{equation}
 \chi^2 = \sum_i^N\frac{(\textrm{data}-\textrm{model})^2}{\sigma^2_i},
\end{equation}
with $i$ labelling the data points (pixels for synchrotron data, point sources for RMs). The variance, $\sigma_i^2$, is of utmost importance and obtaining it is a key accomplishment of this work. 
The variance is primarily \emph{not} experimental or observational uncertainty (which will typically be a negligible contribution), but \emph{astrophysical} variance stemming from random magnetic fields and inhomogeneities in the magnetized ISM. 
In the case of extragalactic RMs there is also a contribution to the variance from Faraday rotation in the intergalactic medium and the source host galaxy. The method we have developed to measure $\sigma^2$ from the data is described in \S\ref{variance_synch} and \S\ref{variance_egs}. This variance map contains valuable information about the random magnetic fields. Indeed, when studying turbulent and small scale magnetic fields this variance map is itself one of the \emph{observables} that a given model of the random field should be able to reproduce.

\subsection{Parameter estimation}\label{paraest}

To minimize $\chi^2$ for a GMF model and find the best fit parameters we implement a Metropolis Markov Chain Monte Carlo algorithm \citep{Metropolis:1953} to explore the likelihood of the observed data. We use the approach outlined in, e.g., \citet{Verde:2003}. For a set of parameters $\vec{p}_j$ we calculate the (unnormalized) likelihood function
$\mathcal{L}(\vec{p}_j)=e^{-\frac{1}{2}\chi^2(\vec{p}_j)}$. We then take a step in parameter space to $\vec{p}_{j+1} = \vec{p}_j+\Delta\vec{p}$, where $\Delta\vec{p}$ is set of Gaussianly distributed random numbers with zero mean and standard deviation $\vec{s}$ (the ``step length'' of the MCMC). If $\mathcal{L}(\vec{p}_{j+1})>\mathcal{L}(\vec{p_j})$ this new set of parameters is accepted, and if $\mathcal{L}(\vec{p}_{j+1})<\mathcal{L}(\vec{p_j})$ the new step is accepted with a probability $\mathcal{P}=\mathcal{L}(\vec{p}_{j+1})/\mathcal{L}(\vec{p}_{j})$. If a step is not accepted, a new $\vec{p}_{j+1}$ is generated, and tested. This process is continued until the parameter space has been sufficiently sampled. To decide when this has happened we use the Gelman-Rubin convergence and mixing statistic, $\hat{R}$ \citep{Gelman:1992}. In the vein of \citet{Verde:2003} we terminate the Markov Chain when the condition $\hat{R}<1.03$ is satisfied for all parameters. To achieve a smooth likelihood surface we run all Markov Chains to at least 50k  steps.

\subsection{Producing simulated observables from models}

We use the \textsc{Hammurabi} code developed by \citet{Waelkens:2008} to simulate full-sky maps of polarized synchrotron emission and rotation measures from 3D models of $n_e$, $n_{cre}$ and $\vec{B}$. The produced polarized synchrotron sky maps (Stokes $Q$ and $U$ parameters) are in HEALPix\footnote{http://healpix.jpl.nasa.gov} format  (an equal-area pixelation scheme of the sphere, developed by \citet{Gorski:2005})  and are calculated taking Faraday depolarization into account. The Galactic contribution to the rotation measure (RM) of extragalactic point sources can also be calculated. For details, see \citet{Waelkens:2008}.

\subsection{Disk - halo separation}\label{diskhalo}

Significant magnetic fields in galactic halos are observed in external galaxies \citep{Beck:2008}. However, the magnetic field in our own Milky Way galaxy has proven very difficult to study, and a long-standing question in Galactic astrophysics is the nature and extent of a magnetic Galactic halo.  For instance it is unclear whether the Galactic halo field may be completely distinct from the magnetic field in the disk, in the sense that their topologies are different. In other words, are the disk and halo fields best modeled by the \emph{same} GMF model (with perhaps different best fit parameters) or distinctly different GMF models? To address this question we separate the sky into a disk part and a halo part, and optimize each GMF model under consideration using only data from one part at a time. 

In the case of polarized synchrotron emission it is very difficult to separate the contribution of the large scale field from that of smaller, more nearby structures simply because the flux from a particular structure scales as $1/r^2$.  Thus we apply a mask on the synchrotron data that covers the disk and clearly defined nearby structures such as the Northern Spur (see section \ref{masking}). Extragalactic rotation measures are not subject to the inverse square effect since the relevent sources are point-like and the RM contribution of some volume of the ISM is independent of its distance from us.

In this paper we will refer to the `disk' and `halo' as distinct regions on the \emph{sky}, where the dividing line between the two is given by the set of directions pointing towards $z=\pm 2$ kpc at Galactocentric radii $R=20$ kpc, i.e., $|b|\lesssim 4\degree$ at $l=0\degree$ and $|b|\lesssim 10\degree$ at $l=180\degree$. This division is plotted in, e.g.,  figure \ref{EGS}.

\section{Data}\label{data}

\subsection{Synchrotron radiation}

In the WMAP K-band (22 GHz) the observed polarized radiation is dominated by Galactic synchrotron emission. In the WMAP five-year data release the observed emission is further analyzed and separated into foreground components caused by synchrotron, dust and free-free emission \citep{Gold:2008}. Throughout our analysis we will use this synchrotron component as our data set for polarized Galactic synchrotron emission. The polarized WMAP component was first included, and analyzed in the context of the GMF, in the three-year data release \citep{wmap_pol:2006}; subsequent GMF modeling of that data set was done by, e.g.,  \citet{Jansson:2007} and \citet{Miville:2008}.

\begin{figure}
\begin{center}
\includegraphics[width=\linewidth]{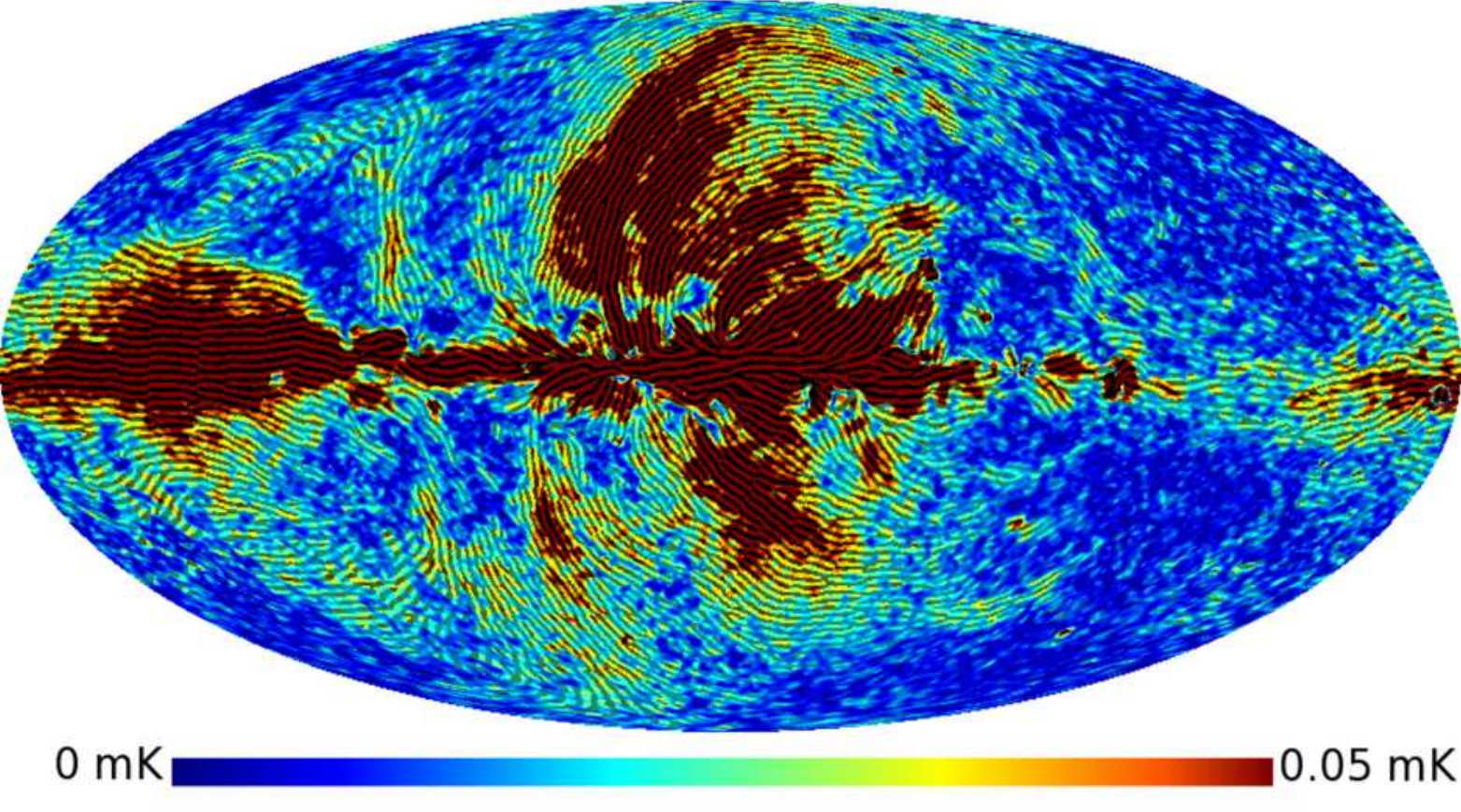}
\end{center}
\caption{Polarized synchrotron intensity (color) in a Mollweide projection with galactic longitude $l=0\degree$ at the center and increasing to the left, overlaid with a texture showing the projected magnetic field directions (i.e., the observed polarization angle rotated by $90\degree$). Image created using the line integral convolution code, ALICE, written by David Larson (private communication).}\label{alice_data}
\end{figure}

The WMAP polarization data is in the form of two HEALPix maps of the Stokes $Q$ and $U$ parameters. The resolution is $\sim$1$\degree$, with about 50k pixels on the sky. To get a more easily interpreted map we can instead plot the polarized intensity $P=\sqrt{Q^2+U^2}$ and  with the polarization angle rotated 90\degree \,to depict the projected magnetic field direction overlaid on the polarized intensity (see figure \ref{alice_data}).

It is clear from figure \ref{alice_data} that emission from the Galactic disk is strong, with polarization angles consistent with a magnetic field parallel to the disk. Also present is the prominent Northern Spur, a highly coherent structure stretching from the disk to high latitudes, and likely the result of the shock between two expanding nearby supernova shells \citep{Wolleben:2007}.

\begin{figure}
\includegraphics[width=\linewidth]{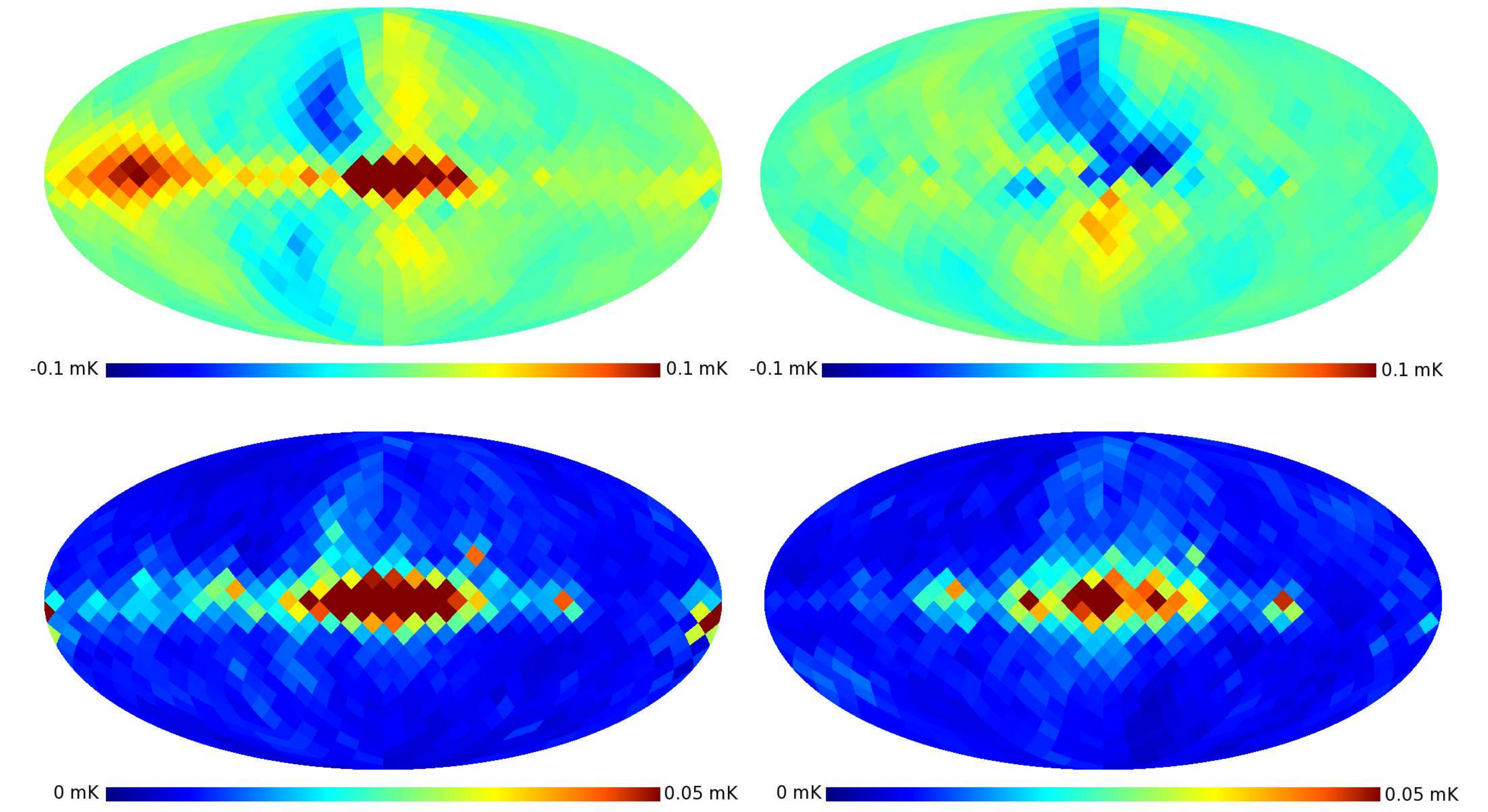}
\caption{\emph{Upper panel:} sky maps of Stokes $Q$ (left) and $U$ (right), in $8\degree$ resolution. \emph{Lower panel:} sky maps of $\sigma_Q$ and $\sigma_U$, respectively. }\label{synch}
\end{figure}

While structures like the Northern Spur can be masked out when studying the large scale GMF, smaller features and irregularities in the polarized emission abound due to the turbulent nature of the magnetized interstellar medium. Since we are interested in the large scale regular magnetic field and want to avoid the random field in this study we smooth the synchrotron data to 8\degree and use maps with 768 pixels, as shown in figure \ref{synch}.

\subsubsection{Estimating the variance}\label{variance_synch}

To calculate $\chi^2$ for the synchrotron data an estimate of the variance is needed that takes into account not just experimental uncertainties but fluctuations in the measured emission due to random magnetic fields. The variance should de-weight the $\chi^2$ of pixels in directions of strong random fields or other localized sources of polarized emission.

We estimate the variance in $Q$ (and similarly in $U$) by 
\begin{equation}
 \sigma^2_{Q, i} = \sum_j^N\frac{(Q_i-q_{j})^2}{N}.
\end{equation}
Here $\sigma^2_{Q, i}$ is the variance of the $i$th pixel in the 8\degree-pixel $Q$ map, whose pixel has the value $Q_i$. The quantity $q_j$ is a pixel value from the 1\degree-pixel $Q$ map, where $j$ enumerates the $N=69$  1\degree-pixels that are within a 4 degree radius of the coordinate on the sky corresponding to the center of the 8\degree-pixel $Q_i$. The lower panels of figure \ref{synch} show sky maps of $\sigma_Q$ and $\sigma_U$.

\subsubsection{Masking}\label{masking}

As pointed out in section \ref{diskhalo}; because the  synchrotron flux from structures such as supernova remnants scale as the inverse square of the distance to the object, flux from nearby structures in the disk can exceed emission caused by the large scale GMF in the diffuse ISM. However not all parts of the disk may  be polluted by bright nearby synchrotron emitters. For example, the large ``Fan region", a strongly polarized part of the sky around $l\sim 120\degree$ to $160\degree$ and $|b|\lesssim 15\degree$ has been argued to be caused by an ordered, large scale ($\sim$kpc) magnetic field oriented transverse to our line-of-sight and with minimal Faraday depolarization \citep{Wolleben:2006}. Other such regions may exist, but it would require more extensive analysis of polarized emission at lower frequencies that probe predominantly the nearby part of the Galaxy (due to Faraday depolarization obscuring more distant emission) in order to use them in the current analysis.

\begin{figure}
\centering
\includegraphics[width=0.7\linewidth]{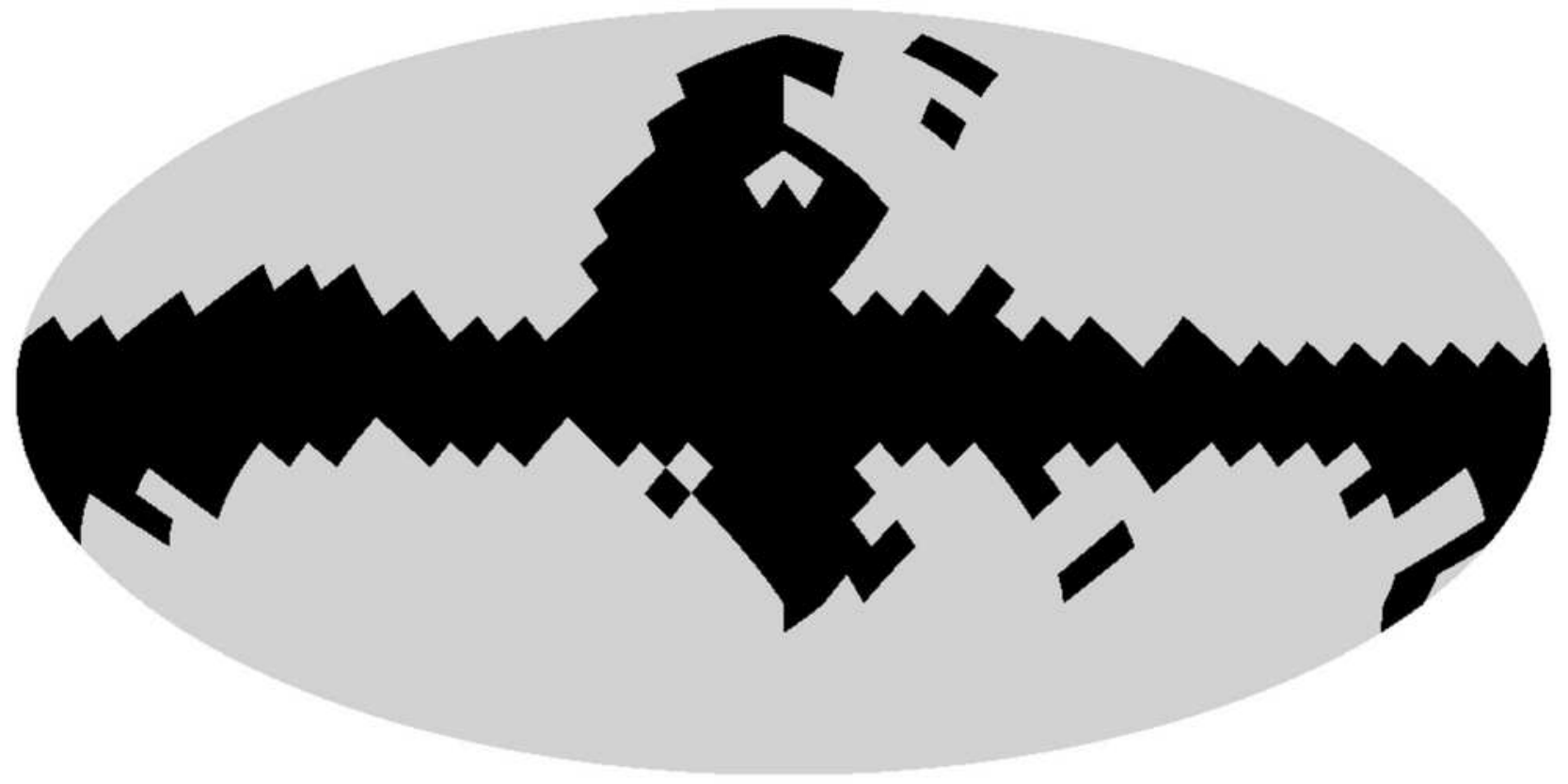}
\caption{The mask used for the polarized synchrotron data. The mask covers 33.5\% of the sky.}\label{mask}
\end{figure}

In this paper we take the simplest approach and mask out the disk and strongly polarized local structures such as the Northern Spur. We use the polarization mask discussed in \citet{Gold:2008}, degraded from $4\degree$ pixels to the $8\degree$ resolution maps we use for the Stokes $Q$ and $U$ data. In cases where an $8\degree$ pixel covers a region of both masked and unmasked $4\degree$ pixels, the larger pixel is made part of the mask. This mask covers 33.5\% of the sky, shown in figure \ref{mask}, and leaves 511 pixels unmasked. Because only a single of these pixels lies inside the `disk', as defined in section \ref{diskhalo}, we  only use polarized synchrotron data when fitting data in the `halo', hence using 510 data points for each of the $Q$ and $U$ maps. 

\subsection{Extragalactic RM sources}\label{egs_rem}

We use 380 RMs of extragalactic sources from the Canadian Galactic Plane Survey (CGPS, \citet{Brown:2003}), 148 RMs from the Southern Galactic Plane Survey (SGPS, \citet{Brown:2007}) and 905 RMs from various other observational efforts 
\citep{Broten:1988, Clegg:1992, Oren:1995, Minter:1996, Gaensler:2001}, for a total of 1433 extragalactic RM sources. The data set contains Galactic longitude and latitude ($l\degree$ and $b\degree$), rotation measure (RM) and observational uncertainty ($\sigma_{obs}$). Typical values of RMs are $\pm$ a few hundred radians/m$^2$ in the disk and  $\pm$ a few tens of radians/m$^2$ at large angles from the disk. The observational uncertainty reported for individual sources tends to be roughly a factor of ten smaller than the typical magnitude of the RM in a given direction.

Some of these 1433 RMs are multiple measurements of the same extended source. Including multiple measurements of one source as if they were different sources would lead us to underestimate the total variance in the rotation measure of that region. To avoid this we round the $l\degree$ and $b\degree$ of each source to one decimal, and if there are multiple RMs with identical coordinates  we replace them with a single RM, whose variance ($\sigma^2_{obs}$) is the mean of the $\sigma^2_{obs}$'s of the individual measurements; the rationale for this is discussed in \S \ref{variance_egs}. This procedure removes 103 `sources', leaving 1330.

\begin{figure}
 \centering
\includegraphics[width=0.8\linewidth]{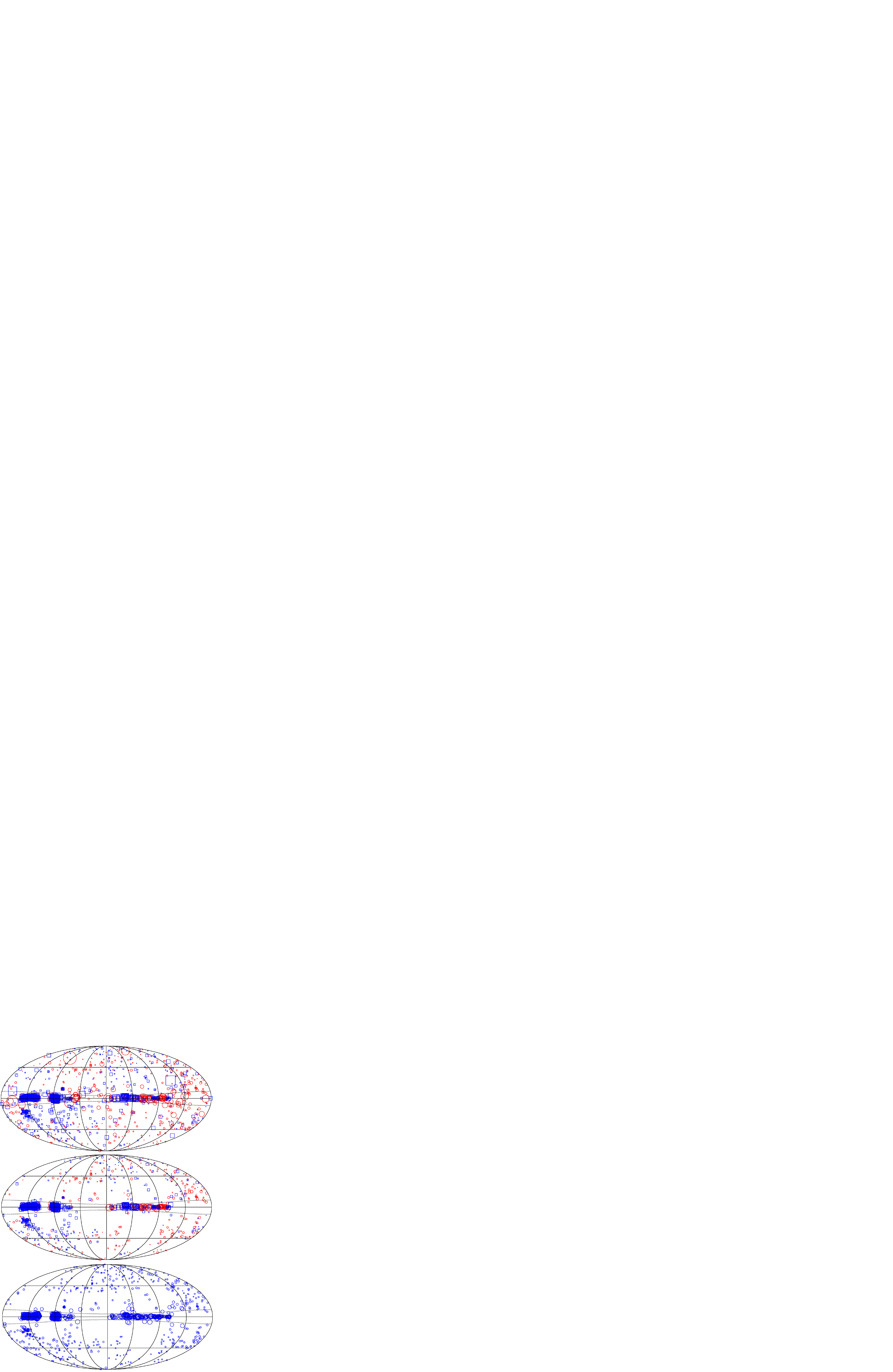}
\caption{\emph{Top:} The full sample of 1433 extragalactic RMs, in Galactic coordinates ($l=0\degree$ at the center and increasing to the left). Blue squares represent negative RM values, corresponding to a magnetic field pointing away from the observer. Red circles show positive RM values. The area of the markers scale as the magnitude of the rotation measure. The black dotted lines marks our division between the disk and halo.
\emph{Middle:} The final sample of 1090 extragalactic RMs (559 in the disk, 531 in the halo) used in the quality-of-fit analysis. 
\emph{Bottom:} The total variance, $\sigma_{EGS}$, of the 1090 RMs used in the data fitting. The area of the markers scale as $\sigma_{EGS}$ and uses the same scale as the RM plots.}\label{EGS}
\end{figure}

\subsubsection{Removing outliers}\label{outliers}

\begin{figure}
\includegraphics[width=\linewidth]{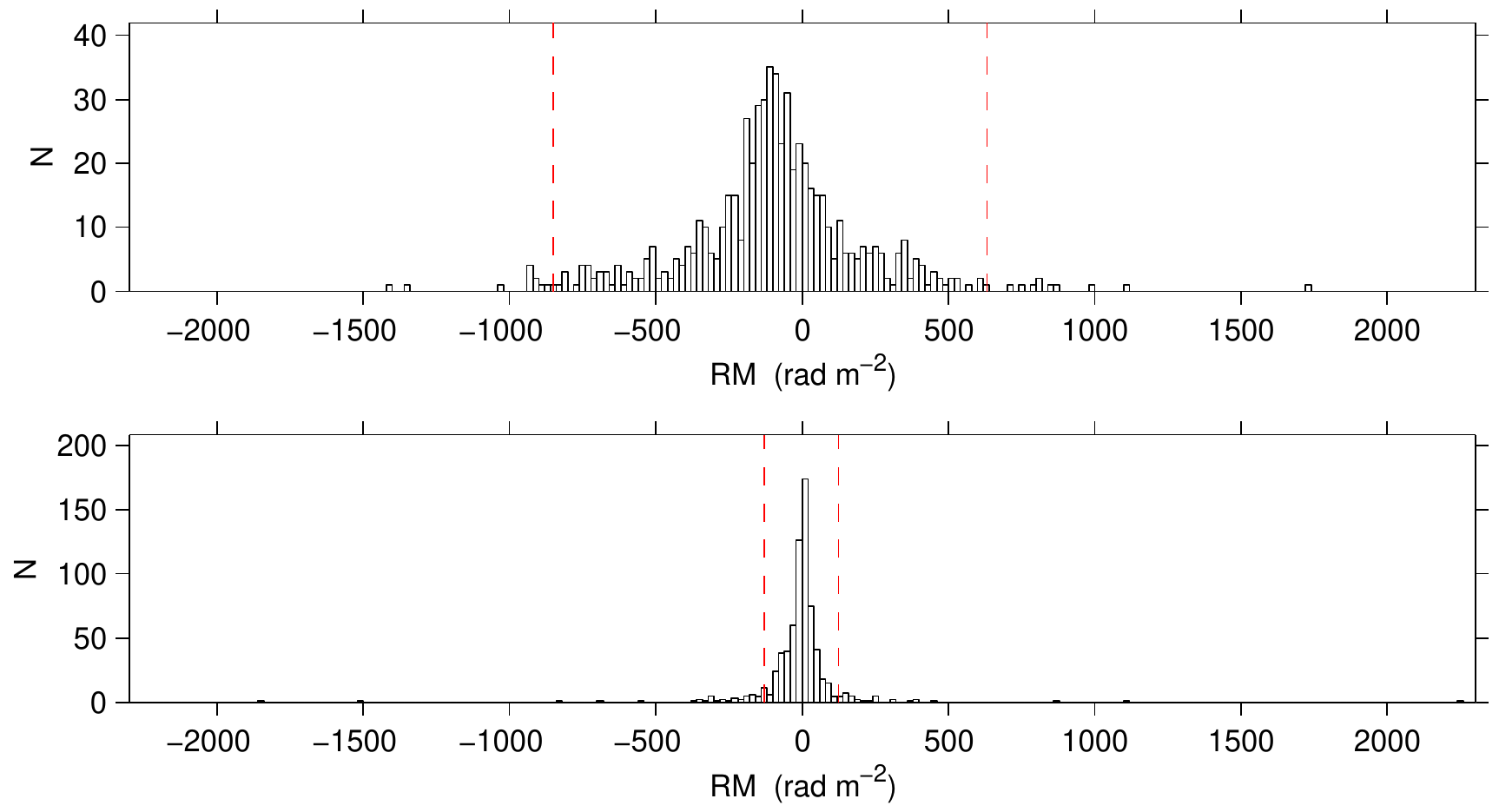}
\caption{Histogram of RMs for the disk region (upper panel) and halo region (lower panel). Dashed lines demarcate RM values excluded as outliers (see section \ref{outliers}).}\label{outlier_hist}
\end{figure}

The measured RM of an extragalactic source is the sum of contributions from the diffuse Galactic interstellar medium (the medium we wish to study), the intergalactic medium and the RM intrinsic to the source. In addition, sightlines that pass through specific structures such as supernova remnants or H II regions can receive very large contributions to their measured RM due the high electron density and possibly strong magnetic fields pervading these structures \citep{Mitra:2003}. Thus it is expected that the variance in the distribution of RM values is significantly larger for sightlines close to the Galactic plane compared to those at large angles to the plane, and this is observed (see figure \ref{outlier_hist}). 

In order to remove the sources whose rotation measure is likely dominated by either highly localized Galactic or purely \emph{non}-Galactic contributions, whose variances are intrinsically different, we divide the RMs into a `disk' component and a `halo' component, in identical fashion to the division described in section \ref{diskhalo}. For each of these we calculate the mean and the standard deviation of the RM values. We use a modified \emph{z-scores} method \citep{Barnett:1984} and exclude any extragalactic sources (EGS) with an RM value three standard deviations away from the mean. These two steps are then repeated until no values of RM are more than three standard deviations away from the mean of the remaining sample. This process removes 21 EGS from the disk and 72 EGS from the halo. 

\subsubsection{Estimating the variance}\label{variance_egs}

As our objective is to study the large scale regular field, the variance $\sigma_{EGS,\,i}^2$ for the $i$th EGS should ideally account for rotation measure contributions of non-Galactic origin and contributions due to small scale random magnetic fields. That is, when these contributions to the rotation measure are large, $\sigma_{EGS,\,i}^2$ should be large, so that the $i$th EGS has a small weight in the calculation of $\chi^2$. There is more than one way to calculate this variance, and we settle on the following scheme to obtain an estimate of the variance:
\begin{equation}
 \sigma_{EGS,\,i}^2 = \sum_j^N\frac{(\textrm{RM}_i-\textrm{RM}_j)^2}{N} + \sigma^2_{obs,\,i},
\end{equation}
where $j$ labels the sources closest to the $i$th source, and $\sigma^2_{obs,\,i}$ is the observational variance. To get a decent estimate of the variance we require a minimum of $N=6$ sources. For a source in the disk, we use all neighboring sources within $\theta_{min}=3\degree$ of the $i$th source. If less than 6 sources are found within $\theta_{min}$ we increase the angular search radius until 6 sources are included or we reach $\theta_{max}=6\degree$. If we still have fewer than 6 sources, we exclude the $i$th source from our sample. For EGS in the halo, where significant changes in the large scale GMF are not expected to occur at as small angular scales as in the disk, we instead use $\theta_{min}=6\degree$ and  $\theta_{max}=12\degree$. 46 sources in the disk and 101 in the halo have an insufficient number of neighboring sources to give a satisfactory estimate of the variance and are excluded. The remaining sample, which will be used in the quality-of-fit calculation, consists of 1090 extragalactic sources and are shown in figure \ref{EGS}; together with the corresponding map of $\sigma_{EGS}$.

Having explained the procedure for measuring $\sigma_{EGS}$, we now return to the issue of how to treat multiple observations of the same source.  The correct procedure depends on the particular situation.  If the source were pointlike and the observations perfectly aligned, then the actual variance in the results of the measurements should in principle be the same as the mean value of the $\sigma^2_{obs}$'s for the set of observations. However if the source is extended and the measurements are not at exactly the same position, their variance probes the $\sigma_{EGS}$ from the foreground turbulent ISM as well as from the RM in the source, which is characteristically much larger than the observational uncertainty in each individual measurement.   Therefore assigning that variance to the combined observation in making the fit would incorrectly reduce its pull in the fit compared to sources with a single measurement.  With the present dataset of RMs, there are only a limited number of cases  of multiple measurements and it is possible to examine them individually.  In only one case is the source extended enough that the  variance in the observations is significantly larger than the mean of the $\sigma^2_{obs}$, and for that case the $\sigma_{EGS}^2$ value we derive is nearly the same with either $\sigma^2_{obs}$ assignment, vindicating {\it a posteriori } our procedure, described in section \ref{egs_rem}.  For future large datasets this issue will require deeper analysis.  If the prescription used here is not adequate, an automatic method will need to be developed to handle all cases, which is applicable for both extremes.

\section{3D models of the magnetized interstellar medium}\label{models}

\subsection{Relativistic electron density}

We use a simple exponential model for the spatial distribution of cosmic ray electrons, 
\begin{equation}
C_{cre}(r,z) = C_{cre,\,0}\,\text{exp}(-r/h_r)\,\text{sech}^2(z/h_z),
\end{equation}
with the scale heights $h_r=5$ kpc and $h_z=1$ kpc (following \citet{wmap_pol:2006}). The normalization factor $C_{cre,\,0}$ is such that for 10 GeV electrons, $C_{cre}(Earth)=6.4\times10^{-5}\,\pcc$, the observed number density of 10 GeV electrons at Earth \citep{Strong:2007}. The number density for other energies is calculated assuming a power law distribution with spectral index $p=-3$. 

\subsection{Thermal electron density}

\begin{figure}
\centering
\includegraphics[width=0.6\linewidth]{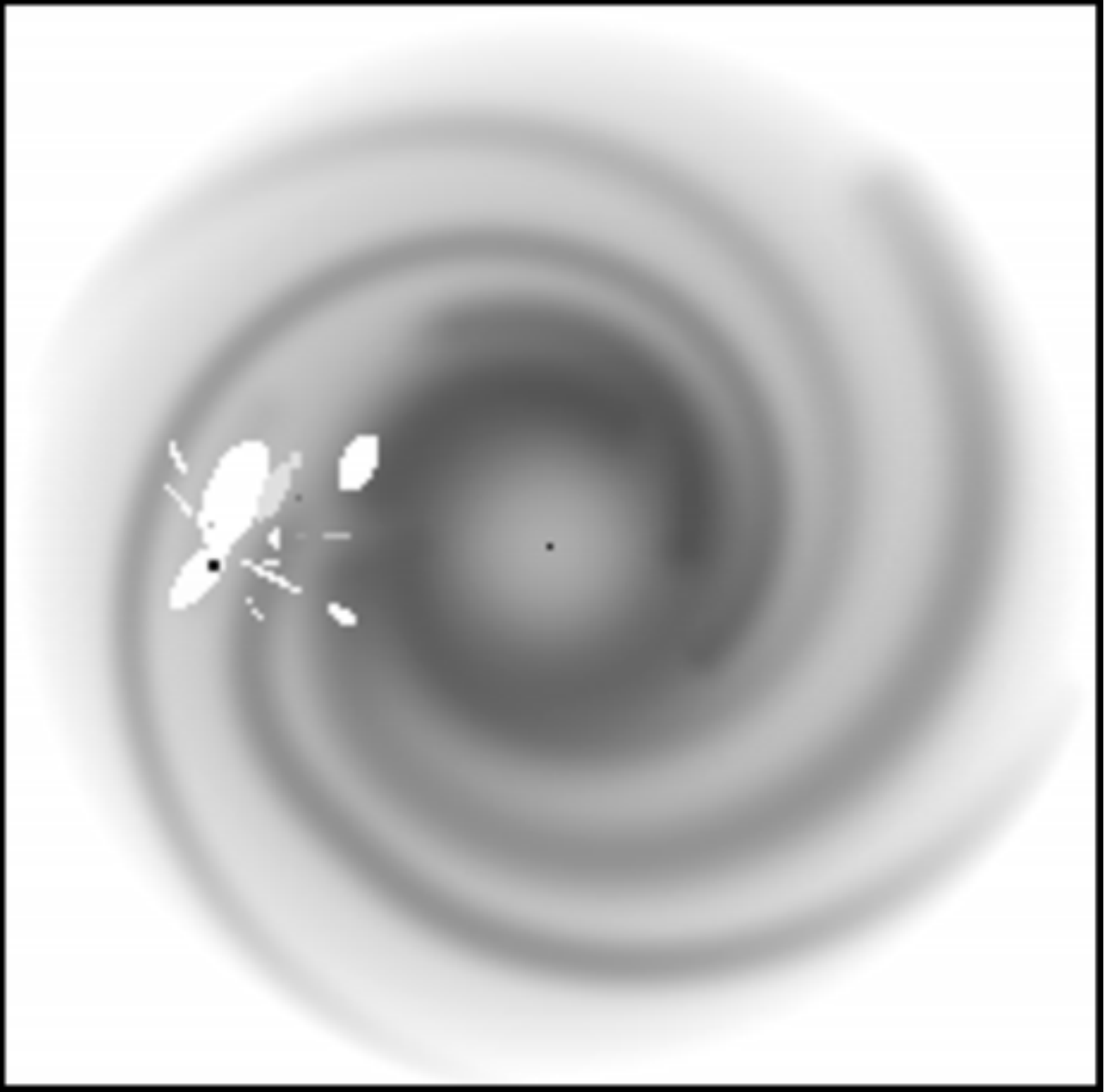}
\caption{The NE2001 thermal electron density model. The figure shows a $30\times30$ kpc Galactic disk at $z=0$. The grayscale is logarithmic. The light colored features in the solar neighborhood are a number of underdense regions. The small dark speck in one of the ellipsoidal underdensities is the nearby Gum nebula and Vela supernova remnant.}\label{NE2001}
\end{figure}

The best available 3D model of the Galactic distribution of thermal electrons is the NE2001 model \citep{Cordes:2002, Cordes:2003}, based on pulsar dispersion measures, measurements of radio-wave scattering, emission measures, and multiple wavelength characterizations of Galactic structure. The model has four different components: a thin disk, a thick disk, spiral arms, and some local over/underdense regions (e.g.,  supernova remnants). The model is shown in figure \ref{NE2001}. The thick disk has a vertical scale height of 0.97 kpc. We adopt NE2001 as our baseline model for the thermal electron density.  We note, however, that the vertical scale height in this model is poorly constrained due to uncertainties in pulsar distances. \citet{Gaensler:2008} estimate the scale height to be 1.83$^{+1.2}_{-2.5}$ kpc, and \citet{Sun:2008} also suggest an increase in the scale height by a factor of two in order to avoid an excessively large magnetic field in the halo. In section \ref{NEscan} we investigate the sensitivity of our conclusions on the choice of electron density scale height.

\subsection{Large scale magnetic field}\label{GMF_models}

Numerous models of the large scale GMF have been proposed in the literature over the last couple of decades. In this paper we test a representative selection of models. 
Most of the considered models were originally proposed to describe either the entire GMF or just the disk field. 

All fields considered here are truncated at Galactocentric radius $R=20$ kpc. For  all models we set the distance from the Sun to the Galactic center to $R_{\sun}=8.5$ kpc. This is the most commonly used distance in the original published forms of the models used in this paper. We note that after a few years of confusion on this topic, the most reliable estimate of this distance is now $8.4\pm0.6$ kpc \citep{Reid:2009}.

\subsubsection{Logarithmic spirals}

This class of models contains the most common models in the literature (see, e.g., \citet{Sofue:1983, Han:1994, Stanev:1997, Harari:1999, Tinyakov:2001}).

We define the radial and azimuthal field components as
\begin{equation}\label{BrBt}
B_r = B(r,\theta, z) \sin p, \hspace{1cm}  B_\theta = -B(r,\theta, z) \cos p,
\end{equation}
where $p$ is the pitch angle of the spiral. The function $B$ is defined as
\begin{equation}\label{Han_B}
 B(r,\theta, z) = -B_0(r) \cos \left(\theta+\beta \ln \frac{r}{r_0}\right) e^{-|z|/z_0},
\end{equation}
with $\beta = 1/\tan p$. The pitch angle $p$ is positive if the clockwise tangent to the spiral is outside a circle of radius $r$ centered on the Galactic center. At the point $(r_0, \theta=180^\circ)$, the field reaches the first maximum in the direction $l=180^\circ$ outside the solar circle. We set the magnetic field amplitude $B_0(r)$ to a constant value, $b_0$, for $r<r_c$, and $B_0(r) \propto 1/r$ for $r>r_c$. With the vertical scale height, $z_0$, the model  has five parameters: $b_0,\, r_c, \,z_0,\,p$ and $r_0$.

The above parametrization is usually named \emph{bisymmetric spiral} (BSS) and exhibits field reversals between magnetic arms (see figure \ref{B_fields}). By taking the absolute value of the cosine in equation (\ref{Han_B}) we get the \emph{disymmetric spiral} (DSS). This latter model is often referred to as an axisymmetric spiral in the literature. However, we prefer to reserve the term `axisymmetric' to mean ``independent of azimuthal angle".  A DSS (BSS) field is a logarithmic spiral field that is symmetric (antisymmetric) under the transformation $\theta \rightarrow \theta + \pi$.

Another distinction that can be made is a model's symmetry properties under $z\rightarrow -z$, i.e., reflection through the disk plane. We denote a field as symmetric (S) with respect to the Galactic plane if $\textbf{B}(r,\theta,-z)=\textbf{B}(r,\theta,z)$, and antisymmetric (A) if $\textbf{B}(r,\theta,-z)=-\textbf{B}(r,\theta,z)$. This notation agrees with, e.g., \citet{Tinyakov:2001}, \citet{Harari:1999}, and \citet{Kachelriess:2005}; however it conflicts with \citet{Stanev:1997} and \citet{Prouza:2003}.

We thus distinguish between four different sub-models of the logarithmic spiral  on the basis of two symmetries; BSS$_S$,   BSS$_A$,  DSS$_S$ and  DSS$_A$.

\subsubsection{Sun et al. logarithmic spiral with ring}

\citet{Sun:2008} test a variety of GMF models for the Galactic disk using extragalactic RMs. The authors find the model best conforming with  data to be an axisymmetric field with a number of field reversals inside the solar circle. Following equation (\ref{BrBt}), Sun et al. defines $B(r,\theta, z) = D_1(r, z)D_2(r)$, with
\begin{equation}
D_1(r, z)= 
\begin{cases} B_0 \,\textrm{exp} \left(-\frac{r-R{\sun}}{R_0} -\frac{|z|}{z_0}\right) & \textrm{if $r>R_c$,}
\\
B_c \,\textrm{exp} \left(-\frac{|z|}{z_0}\right) & \textrm{if $r\le R_c$.}
\end{cases}
\end{equation}
and
\begin{equation}
D_2(r)= 
\begin{cases} +1  & r>7.5 \textrm{ kpc} \\
  -1  & 6 \textrm{ kpc} < r \leq 7.5 \textrm{ kpc} \\
  +1  & 5 \textrm{ kpc} < r \leq 6 \textrm{ kpc} \\
  -1  & r \leq 5 \textrm{ kpc},
\end{cases}
\end{equation}
where $+1$ corresponds to a clockwise magnetic field direction when viewed from the north pole. 

\begin{figure}
\begin{center}
\includegraphics[width=\linewidth]{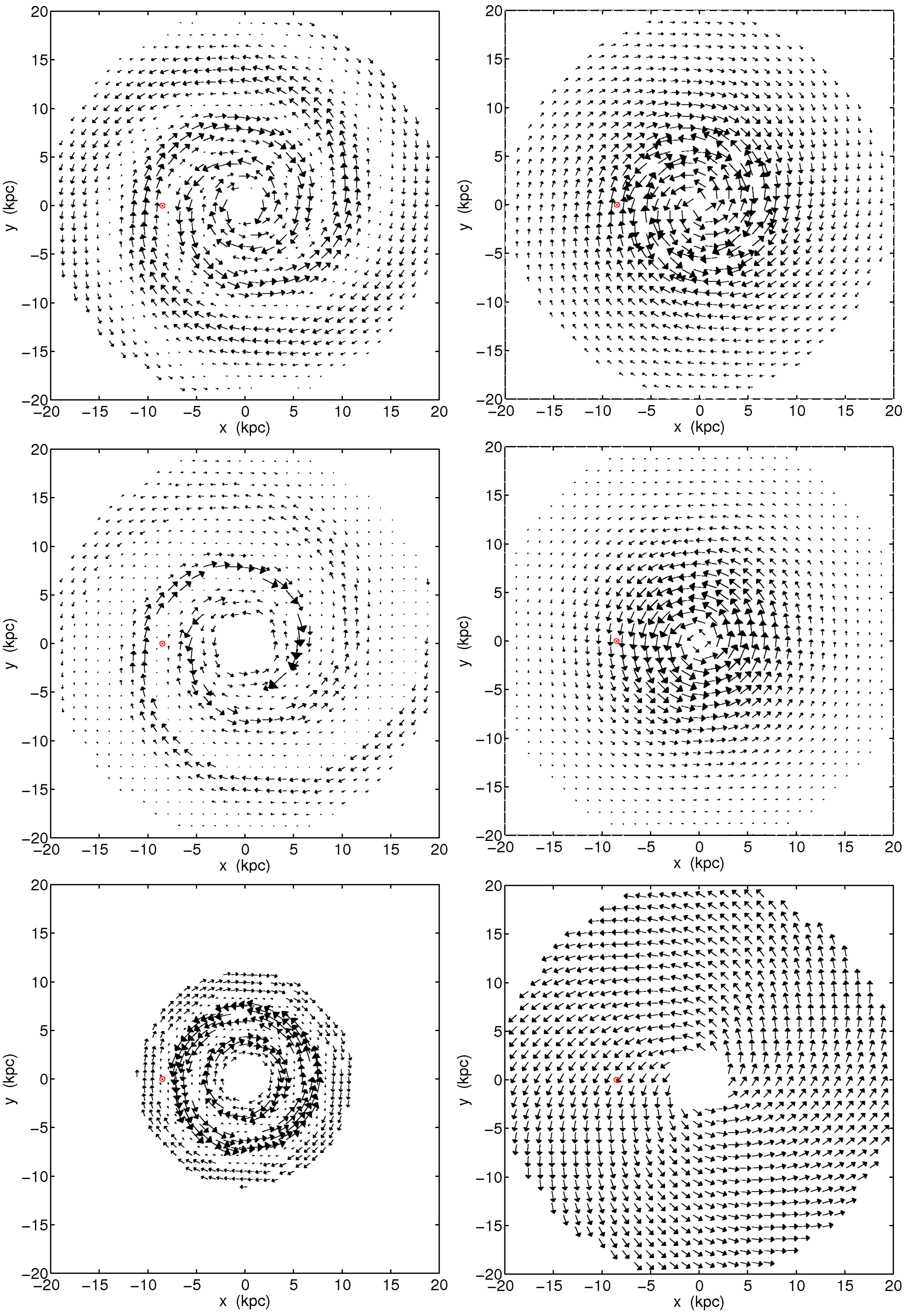}
\end{center}
\caption{\emph{Upper left:} An example of a bisymmetric spiral (BSS) field model for $p=-10\degree$, $r_c=10$ kpc and $r_0=10$ kpc. The corresponding disymmetric spiral (DSS) looks the same except that there is no field reversal between magnetic spiral arms. \emph{Upper right:} Sun08$_\text{D}$, the favored disk model of \citet{Sun:2008}. \emph{Middle left:} The field proposed by \citet{Brown:2007}, based on the NE2001 thermal electron density model. \emph{Middle right:} Halo field model proposed by \citet{Prouza:2003}, and the halo part of the Sun08 composite model. \emph{Lower left:} Toroidal disk field proposed by \citet{Vallee:2008}. \emph{Lower right:} The magnetic field model by the WMAP team \citep{wmap_pol:2006}.}\label{B_fields}
\end{figure}

Since $B(r,\theta, z)$ does not depend on $\theta$, this is an axisymmetric field in the true sense of the word.
Sun et al. take the pitch angle to be the average for the spiral arms in the NE2001 model,  $p=-12\degree$; the other parameters are $B_0=B_c=2\,\muG$, $R_c=5$ kpc, $R_0=10$ kpc and $z_0=1$ kpc. We take these six quantities as parameters to vary in our fit.  This field model (hereafter named Sun08$_\text{D}$) is plotted in figure \ref{B_fields}. 

\citet{Sun:2008} also include a separate halo component, detailed in \S\ref{PS}. In section \ref{results} we consider the full, composite Sun08 model, as well as the disk and halo components separately.

\subsubsection{Prouza-Smida halo field}\label{PS}

The halo component of the Sun08 model is a slightly modified version of the toroidal halo component proposed by \citet{Prouza:2003}. The model (PS03 hereafter) consists of two torus-shaped fields, above and below the Galactic plane, with opposite  field direction (i.e., antisymmetric in $z$). The magnitude of the field is given by
\begin{equation}
B^\text{H}_\phi(r,z) = B_{0}^\text{H}\frac{1}{1+\left(\frac{|z|-z_0^\text{H}}{z_1^\text{H}}\right)^2}\frac{r}{r_0^\text{H}}\text{exp}\left(-\frac{r-r^\text{H}_0}{r_0^\text{H}}\right).
\end{equation}
The parameters used in \citet{Sun:2008} are $B_0^\text{H}=10$ \muG, $r^\text{H}_0=4$ kpc, $z_0^\text{H}=1.5$ kpc, $z_1^\text{H}=z_{1a}^\text{H}=0.2$ kpc for $|z|<z_0^\text{H}$ and $z_1^\text{H}=z_{1b}^\text{H}=0.4$ kpc otherwise. This model is plotted in figure \ref{B_fields}.

\subsubsection{Brown et al. field}

In \citet{Brown:2007} the authors propose a modified logarithmic spiral model\footnote{Details of the model parametrization are not given in \citet{Brown:2007}, but obtained through private communication with the authors, and not reproduced here.} (hereafter Brown07) influenced by the structure of the NE2001 thermal electron density model with the aim to explain the SGPS rotation measure data only (i.e.,  the fourth quadrant region,  $253\degree<l<357\degree$). The model has zero field strength for Galactocentric radii $r<3$ kpc and $r>20$ kpc. Between $3\leq r\leq5$ kpc (the ``molecular ring'') the field is purely toroidal (i.e., zero pitch angle). For $r>5$ kpc, eight magnetic spiral regions with pitch angle $11.5\degree$ are defined with individual field strength $b_j$. The field in the molecular ring and the spiral region corresponding to the Scutum-Crux spiral arm is oriented counterclockwise, and the remaining regions clockwise. The field strength in region $j$ has a radial dependence $|\vec{B}_j|=b_j/r$, with the Galactocentric radius, $r$, in kpc. The vertical extent of the field was not considered, as the model was proposed to explain the measured RMs in the Galactic disk. The model field is shown in figure \ref{B_fields}.

In our analysis we generalize this model by introducing three free parameters: $\alpha$, which scales the overall magnetic field strengths ($b_j$) used in the original model; $r_c$, such that the field strength is constant for $r<r_c$ and $|\vec{B}_j|\propto b_j/r$ for $r>r_c$; and an exponential vertical scale height $z_0$.

\subsubsection{Vall\'{e}e field}

In \citet{Vallee:2008} and \citet{Vallee:2005} the author models the magnetic field in the disk as a perfectly toroidal field consisting of concentric rings of width 1 kpc. The model we  consider here, Vall\'{e}e08, has nine rings between 1 kpc and 10 kpc Galactocentric radii, each with a constant magnetic field strength (see  \citet{Vallee:2008} for details). The field is clockwise as seen from the North Galactic pole, except between 5 kpc and 7 kpc where the field is reversed. In the published model the distance between the Galactic center and the Sun is set to 7.6 kpc. For this reason we rescale the radial location of the boundaries between the magnetic rings by 8.5/7.6. We note that this rescaling still does not allow an entirely fair comparison of the model. The model is shown in figure \ref{B_fields}. The only two parameters we vary in the fit is a single, overall scaling factor for the field strengths and an exponential vertical scale height.

\subsubsection{WMAP field}\label{wmap_field}

In \citet{wmap_pol:2006} the authors cite \citet{Sofue:1986} and \citet{Han_Wielebinski:2002} as reason to model the regular GMF with a bisymmetric spiral arm pattern. They choose the model
\begin{eqnarray}\label{eq:wmap}
 {\bf B}(r,\phi,z) &=& B_0[\sin\psi(r)\cos\chi(z)\hat{r}+ \nonumber \\
& & \cos\psi(r)\cos\chi(z)\hat{\phi}+\sin\chi(z)\hat{z}] 
\end{eqnarray}
where $\psi(r)=\psi_0+\psi_1\ln(r/8 \textrm{ kpc})$ and $\chi(z)=\chi_0\tanh(z/1 \textrm{ kpc})$, and let $r$ range from 3 kpc to 20 kpc.  In \citet{wmap_pol:2006} the distance from the Sun to the center of the Galaxy is taken to be 8 kpc. By fitting model predictions of the  polarization angle $\gamma$ to the WMAP K-band data by using the correlation coefficient $r_c = \cos(2(\gamma_{model}-\gamma_{data}))$, they report a best fit for $\chi_0 = 25^\circ$, $\psi_0=27^\circ$ and $\psi_1=0.9^\circ$. With these parameters $r_{c,\,rms}$=0.76, the rms average over the unmasked sky. This loosely wound spiral is plotted in figure \ref{B_fields}. 

A few comments about the WMAP model: Firstly, \citet{wmap_pol:2006} characterize their model as a bisymmetric spiral, while in fact it is not (for a fixed radius, $|{\bf B}|$ has the same value at all azimuths, and thus is an axisymmetric field). Secondly, using the rms average (as opposed to the arithmetic mean) of $r_c$ leads to an incorrect estimate of the best fit parameters, since, e.g., a pixel with $r_c=-1$ (worst possible fit) is indistinguishable to $r_c=1$ (best possible fit). Finally, equation \ref{eq:wmap} and the above quoted best fit parameters differ from the published quantities due to typos\footnote{Errata: http://lambda.gsfc.nasa.gov/product/map/dr2/pub\_papers/threeyear/polarization/errata.cfm} in \citet{wmap_pol:2006}.

Because the authors  of \citet{wmap_pol:2006} only used the polarization angle in their analysis no constraints on $B_0$ were found. When performing the model fitting with the WMAP field model in this paper we use a four parameter ($B_0,\,\chi_0,\,\psi_0,\,\psi_1$) model. In accordance with \citet{wmap_pol:2006}  no vertical scale height is used.

\subsubsection{Exponential fields}\label{exp_fields}

To investigate the efficacy of specific model features in  lowering the total $\chi^2$ for the optimized model we test a very simple class of GMF models, and iteratively add some complexity. We start with a basic axisymmetric model defined by an overall field strength $B_0$, radial and vertical exponential scale heights $r_0$ and $z_0$, and pitch angle $p$. As in the case of the logarithmic spirals we also test for symmetry/antisymmetry under the transformation $z\rightarrow -z $. We denote these models ``E$_{\textrm{\bf{S}}}$'' and ``E$_{\textrm{\bf{A}}}$''. 

We also test a somewhat more refined model, by introducing a single field reversal. We let $B_0\rightarrow -B_0$ for $r<r_c$, and also allow the pitch angle to be different for $r<r_c$. Again allowing for different symmetry under reflection in the Galactic plane, we label these six-parameter models ``E2$_{\textrm{\bf{S}}}$'' and ``E2$_{\textrm{\bf{A}}}$''.

The antisymmetry with respect to the Galactic plane of the signs of Extragalactic RMs at $|b|\gtrsim15\degree$ has been used to support the idea that the magnetic field in the halo is antisymmetric with respect to reflection through the plane \citep{Han:1997}. However, the antisymmetry of the signs of RMs seem to hold only for the \emph{inner} part of the Galaxy, i.e., $|l|\lesssim100\degree$, which may imply that the GMF in the halo is not globally antisymmetric. To quantify this, we also investigate a third pair of models: ``E2$_{\textrm{\bf{No}}}$'' and ``E2$_{\textrm{\bf{So}}}$''. These are identical to  E2$_{\textrm{\bf{S, A}}}$, except that for  E2$_{\textrm{\bf{No}}}$  the only field reversal occurs at $r<r_c, z>0$ (North), and  at $r<r_c, z<0$  (South) for E2$_{\textrm{\bf{So}}}$.

\section{Results and discussion}\label{results}

\begin{figure}
\includegraphics[width=\linewidth]{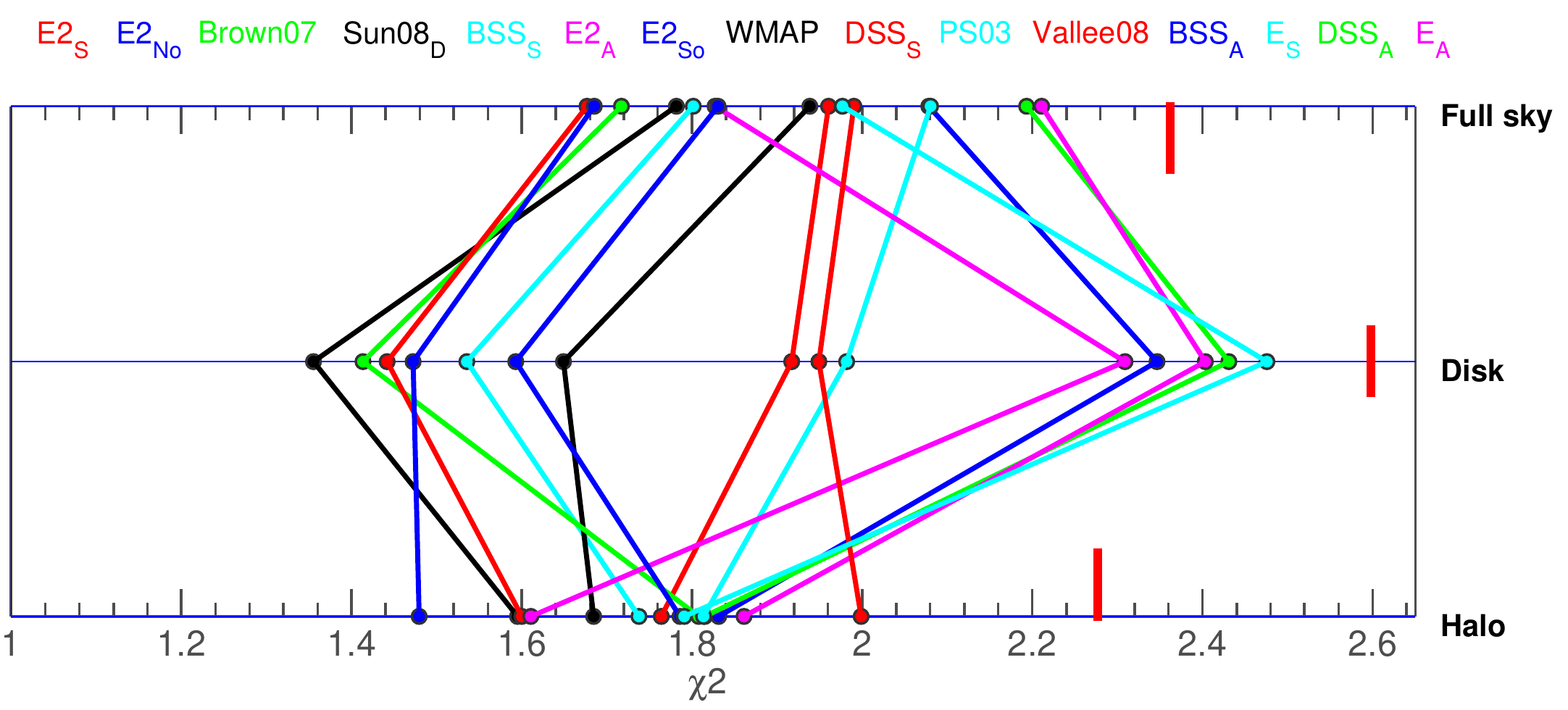}
\caption{The reduced $\chi^2$ for a selection of GMF models. Each model has been optimized to fit the data for the full data set, the disk data, and the halo data, respectively. In each case a separate set of best fit parameters has been obtained. The model names are ordered according to the reduced $\chi^2$  for the full sky data sample. The red bars mark the reduced $\chi^2$ for the null case, $\vec{B}=0$.}\label{chi2_main}
\end{figure}

To quantitatively compare the different GMF models introduced in section \ref{GMF_models} we use  the $\chi^2$ per degree of freedom, also known as the reduced $\chi^2$,
$$
\chi^2_{dof}=\chi_{tot}^2/(N-P),
$$
where $N$ is the number of data points and $P$ is the number of free parameters used in the minimization. For each GMF model we find $\chi_{dof}^2$ for three separate cases: using the complete data set, using only the data in the disk region, and finally using only the halo region data. Each model will thus have three separate sets of best fit parameters. There is no obvious way to best present the results graphically; in figure \ref{chi2_main} we put the reduced  $\chi^2$ of the ``single'' field models (i.e., not the Sun08 composite model, but see \S\ref{fullsun}) for the three cases on three parallel axes, and connect the points for each model. For comparison, we also calculate the reduced $\chi^2$ for the null case ($\vec{B}=0$), which serves as a upper bound on the possible range of $\chi_{dof}^2$, and is marked by a red bar in figure \ref{chi2_main}.

It should be noted that the relative value of the reduced $\chi^2$ for different models is of greater importance than the absolute value for a given model, since the latter is sensitive to somewhat arbitrary choices in the calculation of $\sigma_{EGS}$ and $\sigma_{Q,U}$. We also recall that  in this paper we are not aiming to find \emph{the} best possible model of the GMF, as that would require including random fields and testing a larger set of models. Our purpose is to present a new method of constraining GMF models and draw broad conclusions regarding GMF models currently used in the literature.

From figure \ref{chi2_main}, a few general things  are clear. 
\begin{itemize}
\item There is no overall tendency of field topologies to simultaneously be a good (or bad) fit to both the disk and halo data. This may imply that the Galactic magnetic field in the halo has a topologically distinct structure compared to the disk, i.e., that the GMF is not a ``single'' field with the same form but slightly different parameters in the halo.
\item The spread of minimized $\chi^2_{dof}$ for \emph{disk} data is significant; the models that are antisymmetric under $z\rightarrow-z$ are all strongly disfavored.
\item The spread of $\chi^2_{dof}$ between models when fitting to the halo data alone is much less than when fitting to disk data alone. This is to be expected since polarized synchrotron data is only used in the halo and is identical for $\vec{B}$ and $-\vec{B}$, making for instance a symmetric model indistinguishable from an antisymmetric model. The synchrotron data set also has roughly twice the number of data points as the RM data set in the halo.
\item For the class of axisymmetric models ``E2'', the quality of fit to halo data is remarkably sensitive to the model's specific symmetry with respect to the $z=0$ plane. The case where only the inner region of the Galactic magnetic field is antisymmetric is strongly favored over the completely symmetric or antisymmetric model.

\end{itemize}

\subsection{Fitting models to entire data set}

As noted above, figure \ref{chi2_main} suggests that the magnetic structure of the Galactic disk is significantly different from that of the halo. To test this hypothesis further we investigate the  GMF models with the best performance in the disk and the halo. For these two models, Sun08$_\text{D}$ and E2$_{\textrm{\bf{No}}}$, we note in table \ref{params} the reduced $\chi^2$ for the models when fitted to the region (disk or halo) it performs best in, as well as the reduced $\chi^2$ for the same model when fitted to the other regions (disk, halo, all sky) when the \emph{same} parameters are used as for the best performing region. It is evident that there is a strong tendency for a model to do poorly in regions \emph{not} used in the optimization. Indeed, this effect is so pronounced that when fitted to the disk (halo) data the Sun08$_\text{D}$ (E2$_{\textrm{\bf{No}}}$) model predictions for the halo (disk) is even worse than the null case, $\vec{B}=0$.  We thus conclude that  the disk and halo magnetic fields should be studied separately, and later combined into a complete model of the Galactic magnetic field; as done in, e.g., \citet{Prouza:2003} and \citet{Sun:2008}. Imposing continuity and flux conservation is in general a non-trivial challenge.

\subsubsection{Sun08 composite model}\label{fullsun}

The Sun08 model warrants some additional comments, as it is unique among those considered in this paper in that it has separate components for the disk and the halo. The full model is not included in figure \ref{chi2_main}, since it only makes sense to fit a ``composite" model to the full sky data set. Instead we note the reduced $\chi^2$ for two interesting cases in table \ref{params}. When the full 11 parameter Sun08 model is optimized to the full data set, the lowest $\chi^2$ for any model is achieved. Note however that both the reduced $\chi^2_{disk}$ and reduced  $\chi^2_{halo}$ are 0.2 larger than the lowest $\chi^2$ achieved when fit by separate components (Sun08$_\text{D}$ and E2$_{\textrm{\bf{No}}}$). From figure \ref{chi2_main} it is clear that the PS03 halo component is a relatively poor model of the magnetic field in the halo and is thus forcing the disk component to depart from its preferred parameters to improve the fit in the halo. Because of the large number of parameters (11) we refrain from calculating the confidence levels for the best fit parameters of the full model. Instead we consider the case when the parameters of the disk component are fixed to their optimized values (obtained using disk data only), except for the vertical scale height, $z_0$, which is allowed to vary together with the halo field parameters. The best fit parameters are summarized in table \ref{params}. We note that in combination with the halo field the preferred scale height of the disk component is relatively small ($\sim 0.5$ kpc). The preferred halo field strength is considerable, but as discussed in \citet{Sun:2008} this is most likely due to an underestimation of the thermal electron scale height (see also \S\ref{NEscan} for further discussion).

\subsection{Fitting models to disk data}

\fulltable{\label{params}Best fit parameters and 1$\sigma$ confidence levels for the best fitting models.  The `Region'  column refer to what data set was used to optimize parameters.  The set of reduced $\chi^2$ has been calculated from the model predictions of the listed best fit parameters. 
\linebreak $^\text{a}$ The full Sun08 model with 11 parameters. 
\linebreak $^\text{b}$ The full Sun08 model when keeping the parameters of the disk component (except the vertical scale height) fixed to their best fit values, and varying the halo parameters only.
$^\star$Note:  the likelihood function of $r_c$ for the E2$_{\textrm{\bf{No}}}$ model has a complicated shape, and thus has a less well defined error range (see figure \ref{exp2_n_h_density}). }
\br
Model & \centre{2}{Best fit Parameters} & Region & $\chi^2_{full}$  & $\chi^2_{disk}$ & $\chi^2_{halo}$ \\
\mr
Sun08$_\text{D}$             &  $B_0=1.1 \pm 0.1\,\muG$, & $B_c = 0.16\pm 0.20 \,\muG$   & disk  &  2.27 & 1.35  &  2.61 \\
   &   $p=35 \pm 3\degree$, & $r_c = 5.7\pm 0.2$ kpc  &        &   &   &   \\
   &  $R_0=5.1 \pm 0.9$ kpc &   &   &   &  & \\
 E2$_{\textrm{\bf{No}}}$  &  $B_0=2.3 \pm 0.1\,\muG$,  &   $r_c = 8.72$  kpc$^\star$   &  halo  &   1.86 &    2.94 & 1.48 \\
  &  $p_1=-2 \pm 2\degree$, &  $p_2=-30 \pm 1\degree$  &   &   &   & \\
Sun08$^\text{a}$  &    &    & all  & 1.63  & 1.55  &  1.68 \\
Sun08$^\text{b}$  & $z_0^\text{disk}=0.44 \pm 0.05$ kpc,   & $B^\text{H}_0=4.9\pm 0.7\,\muG$    & all  & 1.68  & 1.42  &  1.78 \\
  & $z_{1a}^\text{H}=0.12\pm 0.05$ kpc  & $z_{1b}^\text{H}=8.5\pm 1.5$ kpc   & & & & \\
  & $r^\text{H}_0=18\pm 4$ kpc & $z_0^\text{H}=1.4\pm 0.1$ kpc   & & & & \\
\br
\endfulltable

\begin{figure}
\begin{center}
\includegraphics[width=0.8\linewidth]{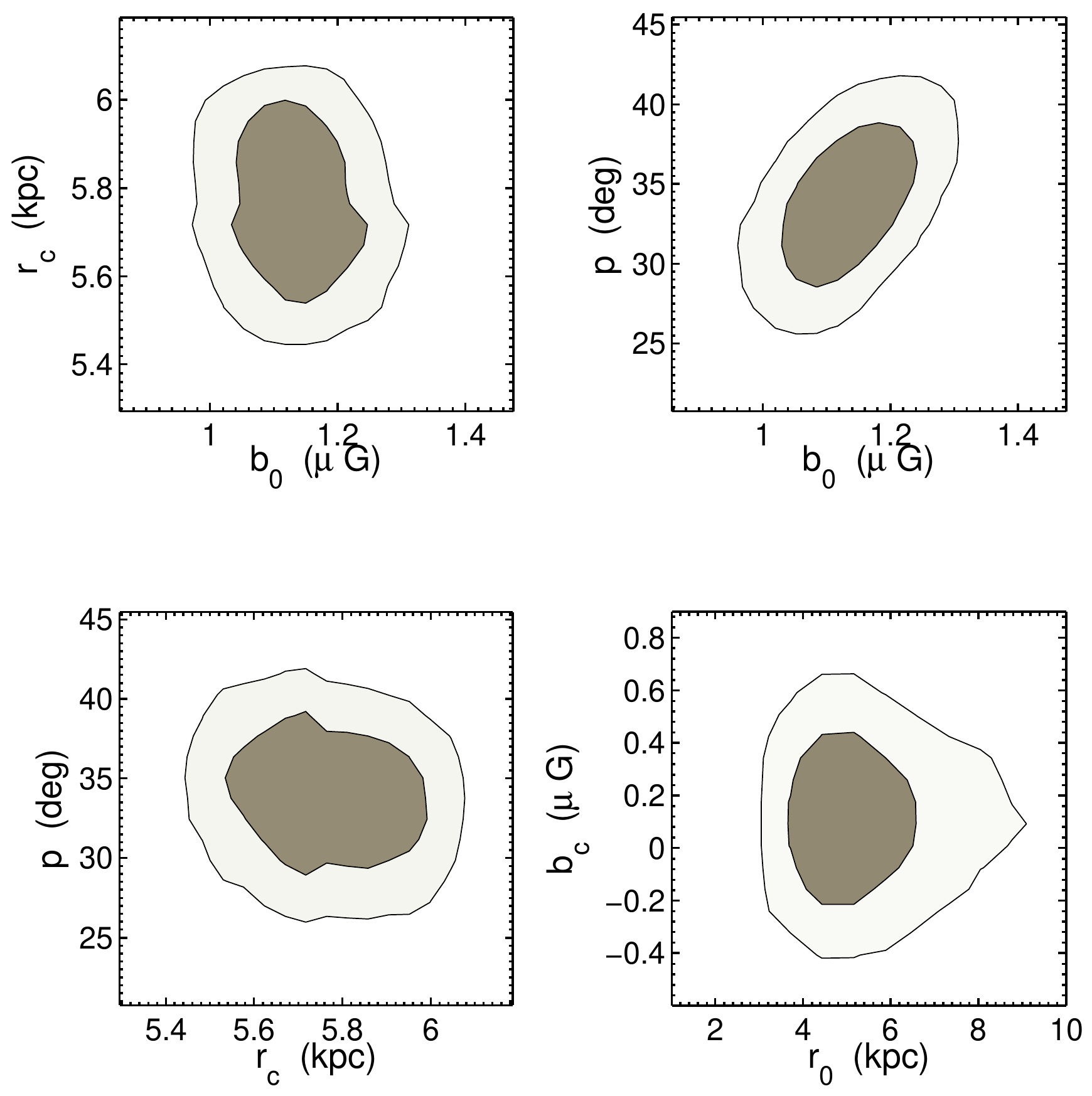}
\end{center}
\caption{The $1\sigma$ and $2\sigma$ confidence levels for the best fit parameters of the Sun08$_\text{D}$ model when fitted to disk data.}\label{sun_density}
\end{figure}

The best fit field model for the disk data is Sun08$_\text{D}$, the axisymmetric field with multiple field reversals  presented in \citet{Sun:2008}. 
The best fit parameters are summarized in table \ref{params} and figure \ref{sun_density}. Note that some of these best fit parameters are very different from the choices in \citet{Sun:2008}. Some differences are  expected since the parameter values were arrived at using slightly different data and a significantly different method of analysis which probes a much larger parameter space. If, e.g., the pitch angle is enforced to the originally proposed value of $p=-12\degree$ the reduced $\chi^2$ optimized for the disk data increases from 1.35 to 1.44, causing the model to lose its lead position.

\begin{figure}
\centering
\includegraphics[width=\linewidth]{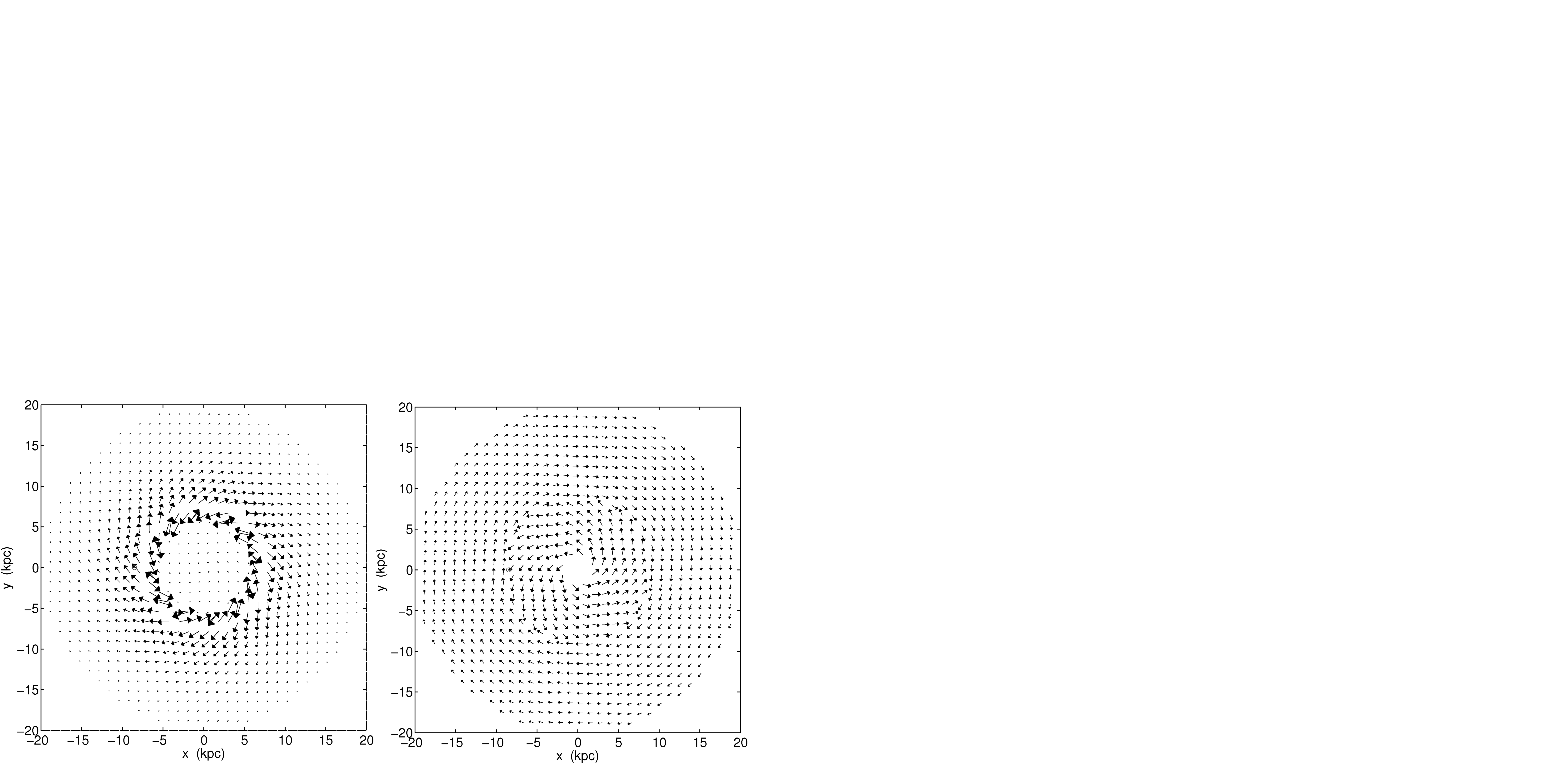}
\caption{\emph{Left:} The best fit Sun08$_\text{D}$-type  model for the magnetic field in disk of the Milky Way. \emph{Right:} The best fit GMF field for the halo data, E2$_{\textrm{\bf{No}}}$. The plot shows the field for $z>0$. In the south ($z<0$),  the field direction in the inner region ($r<8.7$ kpc) is flipped. The relative normalization  is arbitrary.}\label{B_best}
\end{figure}

The runner-up for the best fit disk field is the Brown07 model. It is not surprising that these two models perform the best, since  in addition to the parameters  allowed to vary in the fit (three parameters for Brown07, six for Sun08$_\text{D}$) both models have several fixed parameters that were originally obtained using extragalactic RM data sets similar to the one used in the present analysis. These fixed parameters make a fair comparison of the two models difficult.

\subsection{Fitting models to halo data}

\begin{figure}
\begin{center}
\includegraphics[width=0.8\linewidth]{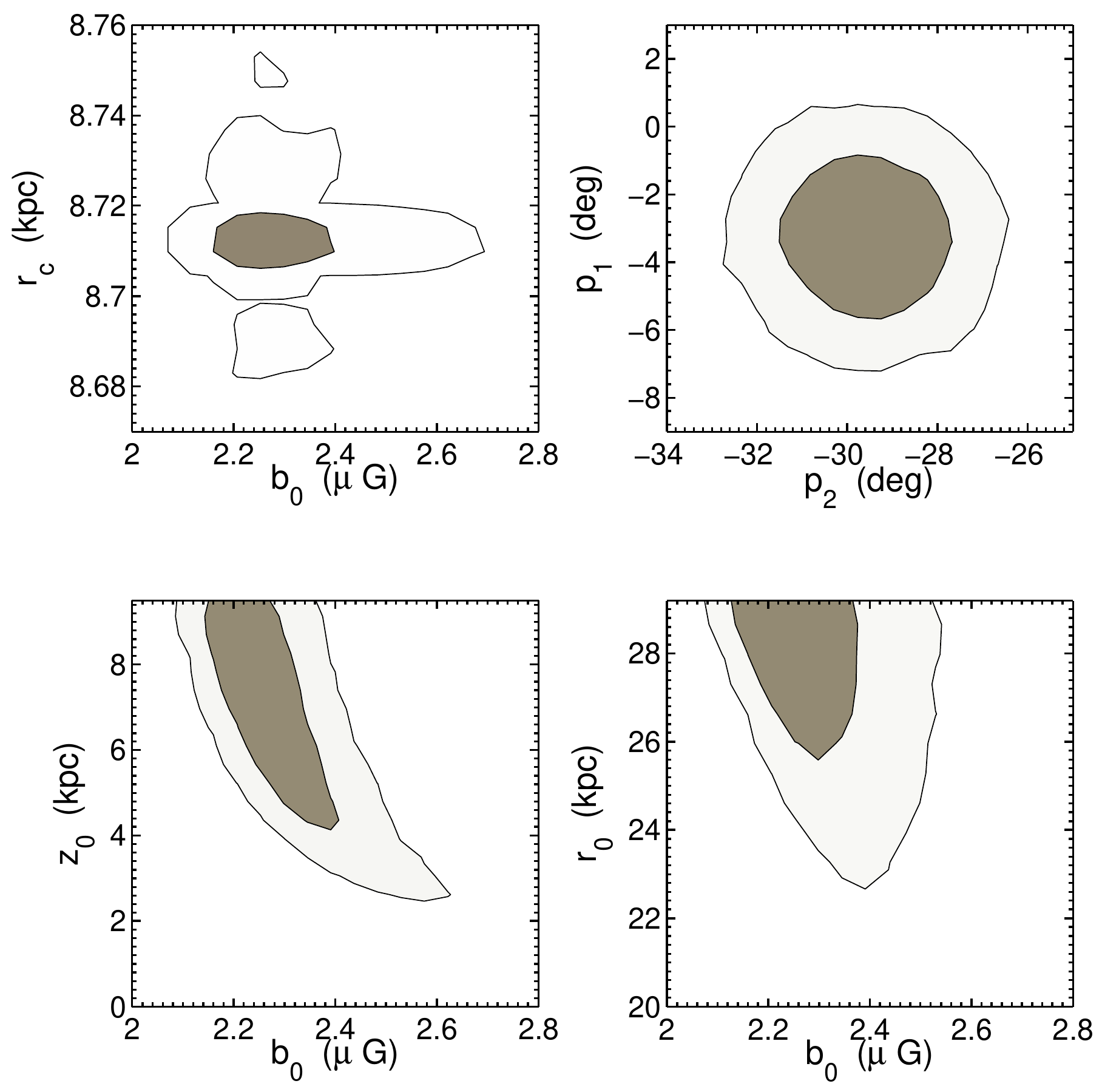}
\end{center}
\caption{The $1\sigma$ and $2\sigma$ confidence levels for the best fit parameters of the E2$_{\textrm{\bf{No}}}$ model when fitted to halo data.}\label{exp2_n_h_density}
\end{figure}

The best fit field model for the halo data is 'E2$_{\textrm{\bf{No}}}$', by a significant margin. Best fit parameters are summarized in table \ref{params} and confidence levels plotted in figure \ref{exp2_n_h_density}, while the field itself is plotted in figure \ref{B_best}. 

Because of the limited radial and vertical extent of the electron distributions, the contribution to the simulated synchrotron emission and rotation measures from regions far from the center of the disk are effectively zero, regardless of the magnetic field in that region. We allow the radial and vertical scale height parameters of the E2$_{\textrm{\bf{No}}}$ magnetic field to vary in the MCMC up to 30 kpc and 10 kpc, respectively. Figure \ref{exp2_n_h_density} makes it clear that very large magnetic scale heights are preferred. If the assumed $n_e$, $n_{cre}$ are correct, we interpret this to mean that a fairly constant field strength is preferred (the spatial extent of which we cannot infer). Alternatively, it can be a sign that the scale heights of the electron distributions are underestimated.

\subsection{Scale heights of electron distributions}\label{scaleheights}

It should be emphasized that synchrotron emission and rotation measures are not observables of the Galactic magnetic field alone, but the magnetized interstellar medium ($\vec{B}, \,n_e,\,n_{cre}$). To properly model the GMF a comprehensive and self-consistent treatment -- where parameters of all three components of the magnetized ISM are optimized simultaneously -- would be desirable, which falls beyond the scope of this paper. 

In this section we briefly investigate how changes in the scale height of electron distributions affect the best fit parameters of some GMF models. This is by no means an exhaustive study, and is mainly intended to test the versatility of our general method of analysis, and to be considered a precursor to future work.

\subsubsection{The vertical scale height of thermal electrons}\label{NEscan}

A result common for all GMF models under consideration is that constraints on their vertical scale height are extremely weak. For the disk, the majority of RMs lie within $|b|<1.5\degree$ and no useful constraint in the vertical magnetic scale height can be obtained. For the halo, constraints on the scale height are also weak. An illustration of this is that a field like the WMAP model which lacks a vertical scale height altogether still can achieve a fairly good $\chi^2$ in the halo. 

The reason the magnetic scale height is poorly constrained is because the observables depend on the product of the magnetic field and an electron density ($n_e$ for RM and $n_{cre}$ for synchrotron emission). The models of these electron density distributions have their own vertical scale heights and thus significantly reduce the RM and synchrotron emission at larger vertical distances from the disk independently of whether a significant magnetic field exists there or not. 

\begin{figure}
\begin{center}
\includegraphics[width=.8\linewidth]{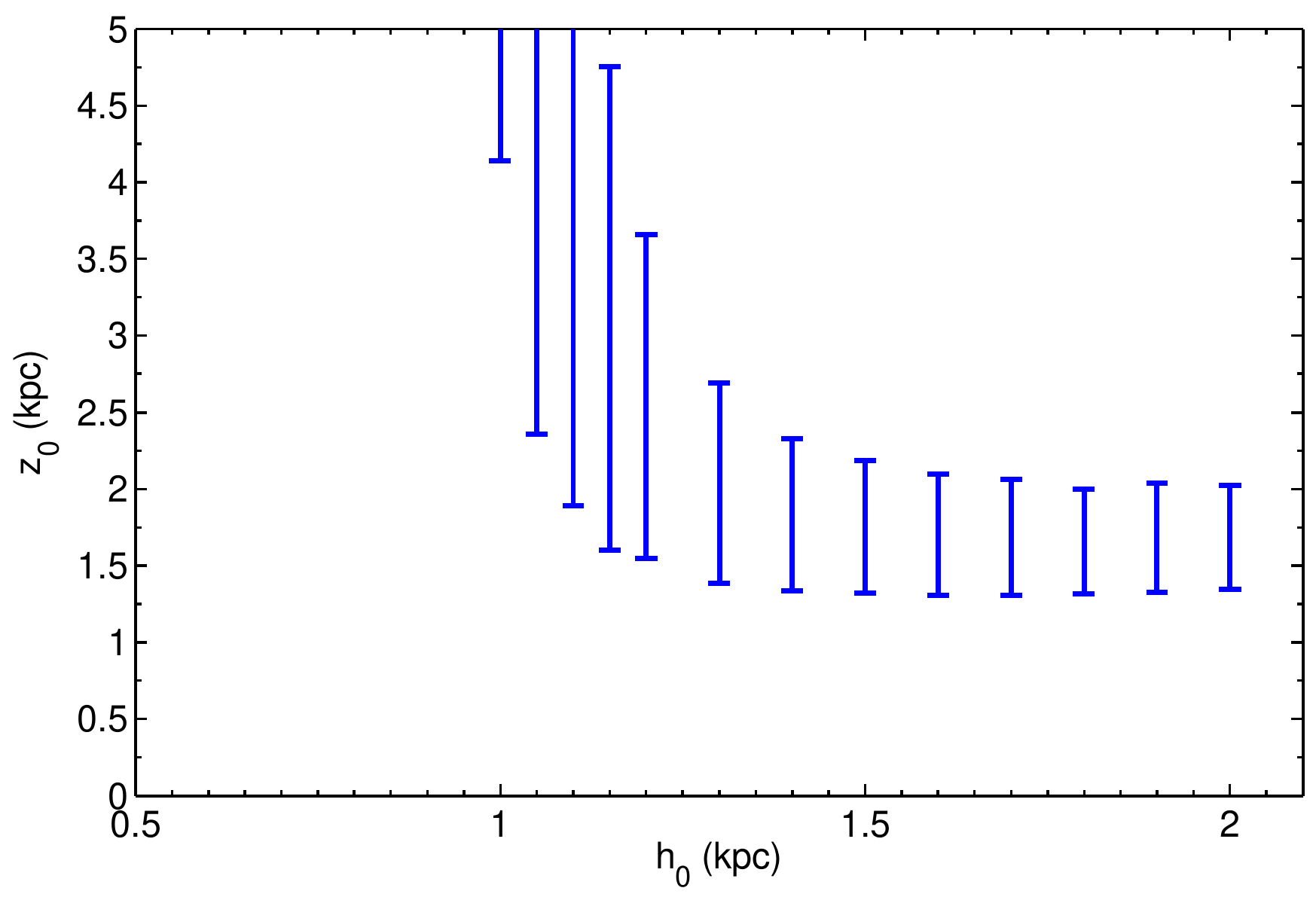}
\end{center}
\caption{The best fit vertical magnetic scale height vs. the vertical scale height of the NE2001 thermal electron density of the thick disk when fit to halo RMs. The E2$_{\textrm{\bf{No}}}$ field model was used, keeping all parameters except $B_0$ and $z_0$ fixed to their best fit values. The error bars contain 90\% of the MCMC sampled points. }\label{fig:NEscan}
\end{figure}

Recent work by \citet{Gaensler:2008} has shown that the vertical scale height of thermal electrons may be significantly underestimated in the NE2001 model, possibly by a factor of two. It is thus important to consider the impact of a change in the thermal electron scale height on the best fit vertical magnetic scale height. To investigate this we modify the vertical scale height, $h_0$, of the thick disk in the \citet{Cordes:2002} NE2001 model (keeping fixed the product of the mid-plane density and the vertical scale height, which is constrained by pulsar DMs), and fit the E2$_{\textrm{\bf{No}}}$ model to the RM and polarized synchrotron data in the halo. All magnetic field parameters except $B_0$ and $z_0$ are kept fixed to their best fit values with the original NE2001 scale height. As seen from the results, which are plotted in figure \ref{fig:NEscan}, the preferred vertical magnetic scale height and its uncertainty decrease as the electron density scale height is increased. 
For a large electron density scale height, $h_0> 1.5$ kpc, the preferred vertical magnetic scale height is $z_0\approx 1.3-2$ kpc. Additionally, if  the vertical scale height of the \emph{relativistic} electron density  tracks $h_0$ and this should be increased as well, the resulting best fit value of $z_0$ would be smaller still.
When in a future analysis both random and regular magnetic fields are considered together, equipartition arguments could be invoked to relate the magnetic field and cosmic ray electron density scale heights.

\subsubsection{Radial scale length of relativistic electrons}

The analysis method introduced in this paper can readily be extended from a formal point of view, to include varying parameters of the electron densities as well as of the GMF models. We test the feasibility of this by letting the galactocentric radial scale length of the relativistic electron model vary (but enforcing the overall normalization such that $n_{cre}(Earth)$ is at its measured value). Redoing the parameter estimation for the halo model E2$_{\textrm{\bf{No}}}$ with this added parameter, we find the new best fit parameters to be within the previously noted confidence levels, and the best fit radial scale length becomes $r_{cre}=6.6\pm 1.5$ kpc, which is close to the fixed value $r_{cre}=5$ kpc.

\subsection{Discussion}\label{discussion}

An obvious but important comment is that a best fit model is not necessarily a `good', or correct, model. And best fit parameters and confidence levels have little relevance if the underlying model is a poor depiction of reality. For example, most of the models under consideration in this paper have a pitch angle as a free parameter, but the best fit pitch angles differ by a factor of a few between models. Hence, if we lack confidence in the veracity of a given model we cannot trust the optimized value of the pitch angle. 

So how do we achieve confidence in a GMF model? More data, in particular RM sources in ``empty'' regions of figure \ref{EGS}, will strongly reduce the number of GMF models that can reproduce the data well. Of special importance are the large gaps of RM data in the disk at $0\degree\lesssim l\lesssim 60\degree$ and $150\degree\lesssim l \lesssim 250 \degree$. At present, a few topologically distinct GMF models considered in this paper reproduce the RM data almost equally well, while at the same time making significantly different predictions for the rotation measure in the regions of the disk lacking data. For example, in the disk the Sun08$_\text{D}$, Brown07 and E2$_{\textrm{\bf{s}}}$ models fit the data best, yet the latter model has only one field reversal and the others three; thus even the sign of RM will differ between the models for part of the disk. Filling these gaps in the data is thus crucial. In addition, more data at high Galactic latitude will enable regions of the sky to be constrained where now there are too few RMs to measure $\sigma$. Pulsar RMs are abundant in the disk, and future work will incorporate this data set as well. In addition, polarized synchrotron data in the disk can be used if the contamination due to local synchrotron features are modeled and subtracted.

A complimentary approach is to introduce  a $\chi^2_{theory}$ term  that incorporates our understanding of physics (e.g., $\nabla \cdot \vec{B}=0$) and information gained through  observations of  magnetic fields in  external galaxies similar to the Milky Way (e.g., that pitch angles only vary within some specific range). Adding $\chi^2_{theory}$  to $\chi^2_{tot}$ would be a first step toward a fully Bayesian analysis. It would act to disfavor unphysical models that, at present, are able to reproduce the current  data well. Similarly, using predictions from a theory of galactic magnetogenesis, such as dynamo theory, could possibly provide powerful constraints on our freedom to build models of the GMF. However, with the genesis of galactic magnetism still not fully understood, we believe it prudent to study the Galactic magnetic field in the most empirical way possible.  Hopefully, a firm observational understanding of the magnetic field of the Milky Way will be a significant aid in resolving the theoretical question of the \emph{formation} of such a magnetic field.

\subsubsection{Model-building}

When building a model of the Galactic magnetic field it is essential to be able to quantify the \emph{improvement} of the model by the inclusion or modification of a particular model feature. This is especially true when models become more complex than what can be described by a handful of parameters, as we have shown in this paper is necessary. The reduced $\chi^2$ provides such a tool as outlined above.  In combination with codes to simulate mock data sets and perform parameter estimation, the reduced $\chi^2$ creates a simulation laboratory where one can easily assess and compare the pull on $\chi^2$ of any proposed model feature.

\section{Summary and conclusion}\label{summary}

In this paper we combined the two main probes of the large scale Galactic magnetic field -- Faraday rotation measures and polarized synchrotron emission -- to test the capability of a variety of 3D GMF models common in the literature to account for the two data sets. We used 1433 extragalactic rotation measures and the polarized synchrotron component of the 22 GHz band in WMAP's five year data release. To avoid polarized local features we applied a mask on the synchrotron data, covering the disk and some prominent structures such as the North Polar Spur. 

We developed estimators of the variance in the two data sets due to turbulent and other small scale or intrinsic effects. Together with a numerical code to simulate mock RM and synchrotron data (the \textsc{Hammurabi} code, by \citet{Waelkens:2008}) this allows us to calculate the total  $\chi^2$ for a given GMF model, and then  find best fit parameters and confidence levels for a given model. This enables us to  quantitatively compare the validity of different models with each other.

From applying our method on a selection of GMF models in the literature we conclude the following:
\begin{itemize}
\item None of  the models under consideration are able to reproduce the data well.
\item The large scale magnetic field in the Galactic disk is fundamentally different than the magnetic field in the halo. 
\item Antisymmetry (with respect to $z\rightarrow -z$) for the magnetic field in the disk is strongly disfavored. For the halo, antisymmetry of the magnetic field in the \emph{inner} part of the Galaxy is substantially preferred  to both the completely symmetric or antisymmetric case.
\end{itemize}

We find the existence of several false maxima in likelihood space to be a generic trait in the models considered. Specifically,  for the best-fit disk model in this work, the true maximum is far removed from the configuration considered in its original form. However, we caution that this optimized configuration should not be trusted as its structure is clearly unphysical. We posit a combination of two explanations for this fact: firstly, the underlying model may not be sufficiently detailed to explain the GMF or may be otherwise incorrect. Secondly, the lack of RM data in large sections of the disk may create both local and global likelihood maxima that would not otherwise exist.

From this and other observations we recommend the following regarding future GMF model-building:
\begin{itemize}
\item It is necessary to probe a large parameter space to avoid false likelihood maxima.
\item A more complete sky coverage of RM data is necessary to derive a trustworthy GMF model.
\item Optimized parameters are sensitive to the underlying electron distributions, and these quantities should be included in the optimization for self-consistency and to achieve correct estimates of confidence levels.
\item When possible, combining separate data sets should be done to maximize the number of data points  in the analysis, to probe orthogonal components of the magnetic field, and to ensure the final model's consistency with the most observables.
\end{itemize}
We also emphasize that by having a well-defined measure of the quality-of-fit, such as provided in this work, the efficacy of any given model feature (i.e., symmetry, extra free parameter, etc.) can be quantified, which is essential when building the next generation of models of the GMF.

In the comparatively near future  the number of measured extragalactic RMs will increase by more than an order of magnitude (see e.g., \cite{Taylor:2009, Gaensler:2004, Gaensler:2009}) and Planck data will become available, which together should allow  large parts of the currently valid model and parameter space to be excluded.

\ack
We thank Marijke Haverkorn for fruitful discussions, and Jo-Anne Brown for helpful discussions and for compiling the list of extragalactic RM sources.
Some of the results in this paper have been derived using the HEALPix \citep{Gorski:2005} package.

\bibliographystyle{jphysicsB}
\bibliography{QU_EGS_061909}

\end{document}